\documentclass[pre,aps,10pt,superscriptaddress,twocolumn,floatfix]{revtex4-1}

\usepackage[nice]{nicefrac}
\usepackage{graphicx,color}
\usepackage{amsmath,amssymb,bm,bbm}
\usepackage{amsmath}
\usepackage{braket}
\usepackage{float}
\usepackage{placeins}

\usepackage{comment}

\usepackage{physics} % for equations
\usepackage[normalem]{ulem} % strikethrough text
\usepackage[T1]{fontenc}
\usepackage[utf8]{inputenc}

\usepackage[plainpages=false,pdfpagelabels,colorlinks=true,linkcolor=red,urlcolor=blue,citecolor=blue,pdftitle={},pdfauthor={},pdfdisplaydoctitle=true,pdfduplex=DuplexFlipLongEdge]{hyperref}

\definecolor{darkred}{rgb}{0.90,0.2,0.2}
\definecolor{darkgreen}{rgb}{0,0.60,.2}
\definecolor{darkblue}{rgb}{0.1,0.3,1}
\definecolor{grey}{cmyk}{0,0,0,0.25}
\definecolor{orange}{cmyk}{0,0.6,0.8,0}

\begin{document}

\title{Sensitivity to perturbations in the three-dimensional Anderson model}

\author{Piotr Tokarczyk}
\affiliation{Institute of Theoretical Physics, Wroclaw University of Science and Technology, 50-370 Wroc{\l}aw, Poland}
\author{Lev Vidmar}
\affiliation{Department of Theoretical Physics, J. Stefan Institute, SI-1000 Ljubljana, Slovenia}
\affiliation{Department of Physics, Faculty of Mathematics and Physics, University of Ljubljana, SI-1000 Ljubljana, Slovenia\looseness=-1}
\author{Anatoli Polkovnikov}
\affiliation{Department of Physics, Boston University, Boston, Massachusetts 02215, USA}
\author{Patrycja  \L yd\.{z}ba}
\affiliation{Institute of Theoretical Physics, Wroclaw University of Science and Technology, 50-370 Wroc{\l}aw, Poland}

\begin{abstract} 
We investigate the fidelity susceptibility, which quantifies the sensitivity of single-particle eigenstates to perturbations, in the three-dimensional Anderson model. 
As a function of disorder strength $W$, it exhibits two distinct peaks. 
The first peak signals a crossover at weak disorder strength from plane-wave states to single-particle quantum chaos, and its position shifts toward $W\to 0$ in the thermodynamic limit. The second peak emerges, to high numerical accuracy, at the critical disorder strength associated with the Anderson localization transition. 
We further show that the divergence of the first peak is maximal, scaling as the square of the inverse frequency cutoff, whereas the divergence of the second peak is submaximal. 
We relate the latter suppression to the fractal structure of single-particle eigenstates at criticality.
We discuss two distinct scenarios that give rise to the peaks in the fidelity susceptibilities.
Moreover, studying the scaling of typical fidelity susceptibilities above the Anderson transition, we find evidence of two distinct regimes of nonergodic behavior.
\end{abstract}

\maketitle

%%%%%%%%%%%%%%%%%%%%%%%%%%%%%%%%%%%%%%%%%
\section{Introduction}

Over the past two decades, significant effort has been devoted to understanding the conditions under which closed quantum systems become ergodic~\cite{Deutsch_PhysRevA_1991,Srednicki_PhysRevE_1994,Rigol2008,D_Alessio_2016,Mori_2018,Deutsch_2018}. 
A promising approach to probing the onset of ergodicity is through the fidelity susceptibility~\cite{PhysRevX.10.041017}, which is closely related to the norm of the adiabatic gauge potential~\cite{Kolodrubetz_2017, Kim_2024}. 
The fidelity susceptibility quantifies the sensitivity of many-body eigenstates to perturbations.

It was proposed that the peak of the fidelity susceptibility encodes information about the crossover from ergodic to nonergodic behavior~\cite{PhysRevX.10.041017}. This idea motivated extensive studies of fidelity susceptibilities in highly excited eigenstates across a variety of quantum systems, including strongly disordered systems~\cite{PhysRevE.104.054105}, systems close to integrability~\cite{PhysRevB.104.L201117, PhysRevB.107.184312, Pozsgay_2024, Rigol_AGP}, random matrix models~\cite{PhysRevB.99.224202, Skvortsov_2022}, toy models of many-body ergodicity breaking~\cite{PhysRevB.111.184203}, Floquet systems~\cite{Bhattacharjee_2024}, fragmented systems~\cite{Lisiecki_2025}, and classical models~\cite{Lim_2024}.

The central question arising from these studies is whether the fidelity susceptibility can be used to quantitatively identify ergodicity-breaking transitions. A natural candidate for such a diagnostic is its peak, which signals maximal sensitivity to perturbations and has also been associated with the point of maximal chaos~\cite{Lim_2024, PhysRevB.111.184203}. However, in most systems studied to date, the nonergodic regime collapses, in the thermodynamic limit, to a single point in parameter space, and the peak position correspondingly drifts toward that point. 
The exception, apart from few-particle systems~\cite{Lim_2024}, is the toy model of many-body ergodicity breaking dubbed quantum sun model~\cite{suntajs_vidmar_22, swietek_hopjan_25}, in which the peak occurs precisely at the ergodicity breaking critical point~\cite{PhysRevB.111.184203}. Nonetheless, a clear demonstration of maximal sensitivity to perturbations of highly excited eigenstates of short-range Hamiltonians, which occurs at a nontrivial parameter value, remains lacking.

Here we address this question in a paradigmatic noninteracting model, the three-dimensional (3D) Anderson model~\cite{PhysRev.109.1492, RevModPhys.80.1355, Suntajs_2021}, which hosts the well-known Anderson localization transition at the critical disorder strength $W_2^*$.
The main result of this work is that the peak of the fidelity susceptibility for several one-body observables emerges, to high numerical accuracy, at $W=W_2^*$.
This indicates that the single-particle eigenstates are maximally sensitive to perturbations at the localization transition.

Moreover, a detailed analysis reveals several additional results. First, we show that for certain observables, such as the sublattice kinetic energy, the fidelity susceptibility also exhibits a peak at weak disorder strength $W_1^*$, where $W_1^*\to 0$ in the thermodynamic limit. 
This peak corresponds to a crossover from localization in quasimomentum space—occurring at the translationally invariant point $W=0$, where eigenstates are plane waves—to single-particle quantum chaos at finite $W>0$. We then analyze the scaling of the fidelity susceptibility peaks with system size, or equivalently with the frequency cutoff $\mu$ associated with its regularization. We find that the peak at $W=W_1^*$ exhibits maximal scaling, diverging as $\mu^{-2}$, whereas the peak at $W=W_2^*$ shows submaximal scaling. 
We attribute this suppression to known properties of the model, including subdiffusive transport~\cite{Ohtsuki_1997, Sierant_2020, hopjan_vidmar_25} and the non-integer fractal dimension of critical eigenstates~\cite{PhysRevLett.83.4590, RevModPhys.80.1355, Rodriguez09, Rodriguez10, PhysRevLett.131.060404, jiricek_hopjan_26} at the Anderson transition. 

Based on our observations, we discuss two distinct scenarios that give rise to the peaks in the fidelity susceptibilities.
The peak at weak disorder emerges due to the relaxation rate given by the Fermi golden rule, which in finite systems may be smaller than the typical level spacing, and the enhancement of matrix elements of observables can be described using fading ergodicity~\cite{Kliczkowski_2024, PhysRevB.111.184203}.
In contrast, the peak at the localization transition point can be explained via the power-law decay of the spectral function, with the exponent given by the wavefunction fractal dimension at the transition point~\cite{PhysRevLett.131.060404}.

Finally, we discuss how differences between the average and typical fidelity susceptibilities may provide additional insight into the properties of the localized phase above the Anderson transition.
In particular, we find evidence of two distinct regimes of nonergodic behavior within the localized phase, and the crossover between them emerges at the disorder strength $W_3^* > W_2^*$.

The paper is organized as follows. In Sec.~\ref{sec2}, we introduce the average and typical fidelity susceptibilities, we explain the importance of regularization, and we discuss different scenarios that give rise to the peak.
The main numerical results for the fidelity susceptibilities in the 3D Anderson model are presented in Sec.~\ref{sec:num}.
In Sec.~\ref{sec:av_versus_typ}, we further analyze the localized phase by comparing  the average and typical fidelity susceptibilities.
We conclude in Sec.~\ref{sec_con}.

%%%%%%%%%%%%%%%%%%%%%%%%%%%%%%%%%%%%%%%%%

%%%%%%%%%%%%%%%%%%%%%%%%%%%%%%%%%%%%%%%%%
\section{Fidelity susceptibility}
\label{sec2}

To probe the sensitivity of single-particle Hamiltonian eigenstates to perturbations, we consider
\begin{equation}
\label{eq_Hlambda}
    \hat{H}_\lambda=\hat{H}_\text{0}+\lambda\hat{O},
\end{equation}
where $\hat{H}_{0}$ is the unperturbed Hamiltonian, $\lambda$ controls the perturbation strength, while $\hat{H}_\lambda |n(\lambda)\rangle = E_n(\lambda)|n(\lambda)\rangle$. To lowest order in $d\lambda$, the overlap between $\ket{n(\lambda+d\lambda)}$ and $\ket{n(\lambda)}$ is given by~\cite{Zanardi_2006,PhysRevE.76.022101}
\begin{equation}
\langle n(\lambda+d\lambda)|n(\lambda)\rangle\approx 1-\chi_n d\lambda^2,
\end{equation}
where the fidelity susceptibility of a single eigenstate is
\begin{equation}
\label{eq:chi_n}
    \chi_n = \sum_{m\neq n}\frac{|\langle n(\lambda)|\hat{O} |m(\lambda)\rangle|^2}{\omega_{nm}^2(\lambda)}\;,
\end{equation}
and $\omega_{nm}(\lambda)=E_n(\lambda)-E_m(\lambda)$ denotes the energy mismatch.
Here we omit the explicit dependence on $\lambda$ and focus on $\lambda=0$.
Moreover, we assume that $\hat{O}$ is local, i.e., it can be written as a sum of operators supported on a vanishing fraction of the system's volume.

We note that one should distinguish two types of perturbations.
The first type of perturbation breaks the underlying localization or integrability of the unperturbed system.
In the 3D Anderson model studied here, this perturbation corresponds to the on-site disorder with amplitude $W$.
The second perturbation is the one introduced by $\hat O$ in Eq.~\eqref{eq_Hlambda} at fixed $W$.
This is what we have in mind when studying sensitivity of the model to perturbations.
However, the choice of $\hat O$ may be nontrivial, and it is one of the goals of this work to explore how it determines the scaling of fidelity susceptibility.

\subsection{Regularization}

The fidelity susceptibility is usually evaluated either in the ground state~\cite{PhysRevLett.98.110601,PhysRevA.77.032111,PhysRevLett.103.170501,PhysRevX.5.031007,PhysRevA.89.033625} or averaged over (a fraction of) energy eigenstates~\cite{Sierant_2019,Lisiecki_2025}. We follow the latter approach. 
The average fidelity susceptibility is generally dominated by contributions with the smallest \(\omega_{nm}\) and hence is not a self-averaging quantity. To address this issue, we define the regularized fidelity susceptibility~\cite{PhysRevX.10.041017,Kim_2024}:
\begin{equation}
\label{eq:chi_av1}
    \chi_\text{av}^r =\frac{1}{D} \sum_{n=1}^D \sum_{m \neq n} \frac{\omega_{nm}^2}{\left(\omega_{nm}^2 + \mu^2\right)^2} |\langle n | \hat{O} | m \rangle|^2,
\end{equation}
where the sum over $n$ runs over energy eigenstates within the energy window of width $\Delta\epsilon$, and $D$ is the number of such states. From now on, we focus on states in the middle of the spectrum and define $Z=\rho(0)$, where $\rho(E=0)$ is the density of single-particle states at energy $E=0$. In this regime, $D$ is typically taken to be proportional to $Z$. Recall that we are interested in the properties of quadratic lattice models, for which $Z\sim V$, where $V$ is the number of lattice sites. The frequency cutoff $\mu$ is chosen such that $\Delta \epsilon \gg \mu > \omega_\text{typ}$, where $\omega_\text{typ}$ denotes the typical level spacing. The physical significance of $\mu$ is revealed when the sums in Eq.~\eqref{eq:chi_av1} are replaced with integrals and a change of variables is performed:
\begin{equation} \label{def_chi_r_av_integral}
     \chi_\text{av}^r=\frac{1}{D}\int dE\mspace{-6mu}\int\limits_{-\infty}^{\infty}\mspace{-6mu}\mspace{-2mu}d\omega\mspace{2mu}\rho\mspace{-4mu}\left(E+\frac{\omega}{2}\right)\rho\mspace{-4mu}\left(E-\frac{\omega}{2}\right)\frac{\omega^2|\langle n | \hat{O} | m \rangle|^2}{(\omega^2+\mu^2)^2},
\end{equation}
where $E=(E_n+E_m)/2$ is the mean energy, $\omega=E_n-E_m$ is the energy mismatch, and the integral over $E$ is taken over the same interval as in Eq.~\eqref{eq:chi_av1}. 

Equation~\eqref{def_chi_r_av_integral} can be further simplified. First, we note that the function $(2\mu/\pi)\omega^2/(\omega^2 + \mu^2)^2$ is peaked at $\omega = \mu$ at small $\mu$. Second, we introduce the coarse-grained function $|f(E,\omega)|^2$, obtained by averaging $\rho(E)|\langle n|\hat{O}|m\rangle|^2$ within a narrow window of $\omega$ of the order of $\mu$. This function is smooth as long as $\mu>\omega_\text{typ}$, irrespective of the properties of the system. However, in the single-particle chaotic systems, which satisfy the single-particle eigenstate thermalization hypothesis (ETH)~\cite{PhysRevB.104.214203, Ul_akar_2022, PhysRevE.109.024102}, the coarse-grained function $|f(E,\omega)|^2$ can be identified with the modulus squared of the envelope function $f(E,\omega)$, which encodes the system's dynamics. The single-particle ETH states that the matrix elements of local observables take the following form:
\begin{equation}
\label{eq:eth}
\langle n| \hat{O} |m\rangle
=\mathcal{O}(E)\delta_{nm}
+\rho(E)^{-1/2}f(E,\omega)R_{nm},
\end{equation}
where $R_{nm}$ are random numbers with zero mean and unit variance, while $\mathcal{O}(E)$ and $f(E,\omega)$ are smooth functions of their arguments. Such systems are sometimes referred to as single-particle ergodic.

%by approximating the off-diagonal matrix elements as $|\langle n | \hat{O} | m \rangle|^2 \approx \rho(E)^{-1} |f(E,\omega)|^2$. This approximation originates from the single-particle eigenstate thermalization hypothesis (ETH), which states that the matrix elements of local observables $\hat{O}$ in single-particle eigenstates satisfy
%\begin{equation}
%\label{eq:eth}
%    \langle n| \hat{O} |m\rangle=\mathcal{O}(E)\delta_{nm}+\rho(E)^{-1/2}f(E,\omega)R_{nm},
%\end{equation}
%where $E=(E_n+E_m)/2$ is the mean energy, $\omega=E_n-E_m$ is the energy mismatch, and $R_{nm}$ are random numbers with zero mean and unit variance. In addition, $\mathcal{O}(E)$ and $f(E,\omega)$ are smooth functions of their arguments. The structure function $\mathcal{O}(E)$ encodes the infinite-time behavior of $\hat{O}(t)$, while the envelope function $f(E,\omega)$ encodes the dynamics of $\hat{O}(t)$. The single-particle ETH has been shown to hold in single-particle chaotic systems, which are therefore sometimes referred to as single-particle ergodic systems~\cite{PhysRevB.104.214203, Ul_akar_2022, PhysRevE.109.024102}.

Since $\rho(E \pm \mu/2) \approx \rho(E)$, one can express Eq.~\eqref{def_chi_r_av_integral} as
\begin{equation}
\label{eq:chi_av_scaling}
    \chi_\text{av}^r \sim \frac{1}{\mu}\int_{-\infty}^{\infty}d\omega \frac{2\mu}{\pi} \frac{\omega^2}{(\omega^2+\mu^2)^2} |f(\omega)|^2 \sim \frac{|f(\mu)|^2}{\mu},
\end{equation}
where $|f(\omega)|^2=1/D\int dE\rho(E)|f(E,\omega)|^2$ corresponds to the spectral function, defined as the Fourier transform of the (connected) autocorrelation function,
\begin{equation}
\label{eq:spectral_rep}
|f(\omega)|^2=\frac{1}{2\pi }\int\limits_{-\infty}^{\infty}\mspace{-6mu}dt e^{i\omega t}C(t)\,,
\end{equation}
with
\begin{equation}
\label{eq:c}
C(t)=\frac{1}{D}\sum_{n=1}^{D}[\langle n|\hat{O}(t)\hat{O}|n\rangle-|\langle n | \hat{O}|n\rangle|^2]\,,
\end{equation}
where $\hat{O}(t)=e^{i\hat{H} t}\hat{O}e^{-i\hat{H} t}$, and we set $\hbar\equiv 1$.

Another way to render the fidelity susceptibility self-averaging --- the approach we will mainly pursue here --- is to consider its typical value rather than its mean:
\begin{equation}
    \chi_{\text{typ}}^{r}=\exp\!\left(D^{-1}\sum_{n}\ln \chi^{r}_{n}\right),
\end{equation}
where $\chi^{r}_{n}$ denotes the fidelity susceptibility of eigenstate $\ket{n}$ with frequency cutoff $\mu$ [which can be obtained from Eq.~\eqref{eq:chi_n} by replacing $\omega_{nm}$ with $(\omega_{nm}^2+\mu^2)/\omega_{nm}$]. 
Its unregularized counterpart satisfies
\begin{equation} \label{def_chi_typ_mu0}
\chi_{\text{typ}}=\chi_{\text{typ}}^{r}(\mu=0)\sim |f(\omega_{\text{typ}})|^{2}/(Z\,\omega_{\text{typ}}^{2})\,,
\end{equation}
because the typical value is attained when the smallest $\omega_{nm}$ is of order $\omega_{\text{typ}}$~\cite{Skvortsov_2022}.

It is worth mentioning that $\chi_\text{av}^r$ from Eq.~\eqref{eq:chi_av1} can be expressed in terms of the Hilbert-Schmidt norm of the adiabatic gauge potential, $\hat{\mathcal{A}}=\frac{1}{2}\int_{-\infty}^{\infty} dt\;\text{sgn}(t)e^{-\mu|t|}\hat{O}(t)$~\cite{PhysRevX.10.041017}. Recently, a different regularization scheme from the one considered in this manuscript has been proposed~\cite{Karve_2025,Author2026}. Specifically, $\mu^{-1}$ is replaced with a hard time cutoff $T$, for which $\hat{\mathcal{A}}=\frac{1}{2}\int_{-T}^{T} dt\;\text{sgn}(t)\hat{O}(t)$. This leads to a very similar expression to the one for $\chi_\text{av}^r$ in Eq.~\eqref{eq:chi_av_scaling}, but with $(2\mu/\pi)\omega^2/(\omega^2+\mu^2)^2$, which is peaked at $\omega=\mu$, replaced by $(\pi T)^{-1}(1-\cos(\omega T))^2/\omega^2$, which is peaked close to $\omega=\pi T^{-1}$. Therefore, the two regularization schemes exhibit similar behavior when $T\sim\mu^{-1}$. 

The above discussion demonstrates that $|f(\mu)|^2$ and, therefore, $\chi_\text{av}^r$ and $\chi_\text{typ}^r$, probe the system's dynamics up to times $t\sim\mu^{-1}$. This also implies that the systems exhibiting larger $\chi_\text{typ}^r$ are not only more unstable to small changes in $\lambda$, but are also characterized by a longer relaxation time of $\hat{O}$. In contrast, the unregularized version, $\chi_\text{typ}$, probes the system's dynamics up to the longest physical time $t=\tau_\text{H} = \omega_\text{typ}^{-1}$, known as the Heisenberg time.

We note that although $\chi^{r}_{\text{typ}}$ is well defined even in the limit $\mu\!\to\!0$, it is often convenient to retain the dependence on $\mu$. Generally, the limits $\mu \rightarrow 0$ and $V \rightarrow \infty$ are not expected to commute, just as the limits $t \rightarrow \infty$ and $V \rightarrow \infty$ are known not to. Moreover, the Heisenberg time is often inaccessible in experimental setups, so that it can be treated as effectively infinite. For example, note that in classical systems the Hilbert-space dimension is effectively infinite, while the fidelity approach remains well suited to probe chaos~\cite{Lim_2024, Karve_2025, karve2025diffusionsignaturechaos}. In such situations it is preferable to work in the thermodynamic limit and analyze how the fidelity susceptibility scales with $\mu$, rather than to study its finite-size scaling. Although in our system the full Hilbert space is accessible, we retain the finite-$\mu$ analysis to highlight that it provides useful insight without the need to reach the Heisenberg time.

%%%%%%%%%%%%%%%%%%%%%%%%%%%%%%%%%%%%%%%%%

\subsection{Scaling of the peak of fidelity susceptibility} \label{sec:peak_scaling}

Here we introduce two scenarios, referred to as scenario 1 and scenario 2, which address the origin of peaks in $\chi^r_\text{av/typ}$.
While this discussion is general and inspired by some recent insights in the literature, we will apply the results of this section to explain some of the numerical results in Sec.~\ref{sec:num}.

In scenario 1, we assume the Lorentzian broadening of the peak in the spectral function of an observable $\hat O$,
\begin{equation} \label{def_lorentzian}
    |f(\omega)|^2\sim {1\over \pi} {\Gamma\over \Gamma^2+\omega^2},
\end{equation}
where the width $\Gamma$, corresponding to the relaxation rate, gradually decreases when approaching the transition point from the ergodic side.
This is typically the case in systems where $\Gamma$ is given by the Fermi golden rule.
Examples of such behavior have been recently reported in systems that exhibit fading ergodicity~\cite{Kliczkowski_2024, swietek_kliczkowski_26} and in certain classical systems~\cite{kim2025confineddeconfinedchaosclassical}.
In the context of integrability-breaking transitions, the observable $\hat{O}$ for which Eq.~\eqref{def_lorentzian} can apply has to have a non-zero projection onto the integrals of motion at the integrable point.
Using Eq.~\eqref{def_lorentzian}, we find 
\begin{equation}
\label{eq:chi_Gamma}
\chi^r_\text{av}(\mu,\Gamma)\sim\int_{-\infty}^{\infty}\frac{d\omega}{\pi}\,
\frac{\Gamma}{\Gamma^2+\omega^2}\,
\frac{\omega^2}{(\omega^2+\mu^2)^2}
=\frac{\Gamma}{2\,\mu\,(\mu+\Gamma)^2},
\end{equation}
and a similar scaling is anticipated for $\chi^r_{\rm typ}(\mu,\Gamma)$. 
Hence, at a fixed $\mu$, it scales as 
$\chi^r_\text{av}(\mu, \Gamma) \sim \frac{\Gamma}{2 \mu^3}$ for $\Gamma \ll \mu$, i.e., it increases with $\Gamma$. 
It reaches a maximum of $\chi^r_\text{av}(\mu, \mu) = \frac{1}{8 \mu^2}$ at $\Gamma = \mu$ and then it decreases as $\chi_\text{av}^r(\mu, \Gamma) \sim \frac{1}{2 \mu \Gamma}$ for $\Gamma > \mu$. At a fixed $\Gamma$, $\chi_\text{av}^r(\mu,\Gamma)$ is a monotonically decreasing function of $\mu$. Therefore, it diverges in the limit $\mu\to 0$~\cite{Sierant_2019,PhysRevE.104.054105}. At the same time, the typical susceptibility saturates to
\begin{equation}
\chi^r_{\rm typ}(\mu=0,\Gamma)\equiv \chi_{\rm typ}(\Gamma)\sim \frac{Z}{\Gamma}\,.
\end{equation}
In the fading ergodicity~\cite{Kliczkowski_2024}, the ergodicity breaking critical point is characterized by $\Gamma\propto Z^{-1}$, and hence the fidelity susceptibility saturates its upper bound, i.e., $\chi_{\rm typ} \propto Z^2$~\cite{PhysRevB.111.184203}. 

Scenario 2 relies on the slowing down of the relaxation of observables towards equilibrium, which is encoded in the long time asymptotic behavior of the autocorrelation function, i.e., $\langle \hat{O}(t)\hat{O} \rangle=\frac{1}{Z}\sum_{n}\langle n |\hat{O}(t)\hat{O} | n \rangle$. This scenario can be observed near the localization transitions in disordered systems~\cite{Skvortsov_2022}, the continuous phase transitions characterized by the critical slowing down of relaxation dynamics~\cite{HohenbergHalperin1977}, or in the low-dimensional classical nonlinear systems~\cite{Lim_2024}. The autocorrelation function then decays as a power-law in time, i.e., $\langle \hat{O}(t)\hat{O} \rangle \sim t^{-\bar a}$, with $\bar a\in[0,1]$ at the transition point~\cite{PhysRevE.104.054105, Skvortsov_2022, Lim_2024}. In this scenario, the maximum of the fidelity susceptibility is reached at the transition point with $\chi_\text{typ}^r
\sim \mu^{-(2 - \bar a)}$ leading to
\begin{equation} \label{def_bar_a}
{\rm max}_{\mu}[\chi^r_\text{typ}]={\rm max}[\chi_\text{typ}] \sim Z^{2 - \bar a}\,. 
\end{equation}
Only in the limit $\bar a \to 0$, this maximum does reach the upper bound, $Z^2$, which corresponds to the regime of logarithmic relaxation in time or the so-called $1/f$-noise~\cite{Paladino2014fNoiseReview}.

In Appendix~\ref{sec:app_phenomenological}, we provide a phenomenological model in which both scenarios arise as distinct limiting cases. It is based on the idea of Drude peak broadening as a result of local integrals of motion acquiring finite relaxation rates. Although the model is phenomenological, we believe it correctly captures the origin of the peak observed in the fidelity susceptibilities during integrability-breaking transitions.

\subsection{Many-particle states of quadratic models}

In passing, we emphasize that all numerical calculations presented in the manuscript concern single-particle eigenstates, as numerical calculations involving many-body eigenstates are computationally more challenging. Nevertheless, we can argue that the behavior of fidelity susceptibilities is expected to be similar for both single-particle and many-body eigenstates of quadratic models. 
To this end, let us write the fidelity susceptibility of a many-body eigenstate~$\ket{\mathcal{F}_a}$ with a corresponding many-body energy~$\mathcal{E}_a$ as 
\begin{equation}
    \chi_{a}=\sum_{b\neq a}\frac{|\langle \mathcal{F}_a|\hat{O}|\mathcal{F}_b\rangle|^2}{(\mathcal{E}_a-\mathcal{E}_b)^2}.
\end{equation}
Additionally, let us express a one-body operator as 
\begin{equation}
    \hat{O}=\sum_{n,m}O_{nm}\hat{f}_n^\dagger \hat{f}_m,
\end{equation}
where $O_{nm}$ are the matrix elements in the single-particle eigenbasis, while $\hat{f}_n^\dagger$ ($\hat{f}_n$) creates (annihilates) a spinless fermion in a single-particle eigenstate~$\ket{n}$.
Assuming a fixed particle number, and taking into account that many-body eigenstates are configurations of occupied single-particle eigenstates for quadratic Hamiltonians, we note that $|\langle \mathcal{F}_a|\hat{O}|\mathcal{F}_b\rangle|=|\sum_{n,m}O_{nm}\langle \mathcal{F}_a|\hat{f}_n^\dagger \hat{f}_m|\mathcal{F}_b\rangle|$ is nonzero only for pairs of many-body eigenstates that differ by a single-particle excitation. For such pairs, the matrix element reduces to $|\langle \mathcal{F}_a|\hat{O}|\mathcal{F}_b\rangle|=|O_{nm}|$, where $\ket{n}$ ($\ket{m}$) is an occupied and $\ket{m}$ ($\ket{n})$ is an unoccupied single-particle eigenstate in $\ket{\mathcal{F}_a}$ ($\ket{\mathcal{F}_b}$). Moreover, the corresponding many-body energy difference is $\mathcal{E}_a-\mathcal{E}_b=E_n-E_m=\omega_{nm}$. 
Therefore, the fidelity susceptibility of a many-body eigenstate simplifies to 
\begin{equation}
\label{eq:final}
    %\chi_a=\sum_{n\in\mathrm{occ}}\sum_{m\in\mathrm{unocc}}\frac{|O_{mn}|^2}{\omega_{nm}^2}
    \chi_a=\sum_{n,m=1}^{D}\frac{|O_{mn}|^2}{\omega_{nm}^2}p_n(1-p_m)\approx \frac{1}{4}\sum_{n,m=1}^{D}\frac{|O_{mn}|^2}{\omega_{nm}^2},
\end{equation}
where $p_n$ is the occupation of the single-particle states $\ket{n}$ in the many-body state~$\ket{\mathcal{F}_a}$. To establish the last expression, we have considered the half-filling case, for which $p_n\approx 1/2$. Comparison of $\chi_a$ from Eq.~\eqref{eq:final} with $\chi_n$ from Eq.~\eqref{eq:chi_n}, justifies the expectation of similar behavior of fidelity susceptibilities for both single-particle and many-body eigenstates of quadratic models, such as the Anderson model studied below. Nevertheless, we expect the eigenstate-to-eigenstate fluctuations of the fidelity susceptibility at finite $\mu$ to be suppressed in a many-particle system. They can become important again when $\mu$ is of the order of the many-particle level spacing.\\

%%%%%%%%%%%%%%%%%%%%%%%%%%%%%%%%%%%%%%%%%
\section{Numerical results for \\Anderson model}
\label{sec:num}

We now study the fidelity susceptibility in the 3D Anderson model, which is the central part of this work. 

%\subsection{Model and perturbations}

The Hamiltonian of the 3D Anderson model on a cubic lattice can be written as~\cite{PhysRev.109.1492}
\begin{equation}
\label{eq_HA}
    \hat{H}_\text{A}=-\sum_{\langle i,j \rangle} \hat{c}_{i}^\dagger \hat{c}_{j}+\sum_{i=1}^{V} \varepsilon_{i} \hat{c}_{i}^\dagger \hat{c}_{i},
\end{equation}
where $\hat{c}_{i}^\dagger$ ($\hat{c}_{i}$) creates (annihilates) a spinless fermion at site $i$. The first term in Eq.~(\ref{eq_HA}) describes the hopping between the nearest neighbors, $\langle i,j \rangle$. The second term describes uncorrelated disorder, with $\varepsilon_i$ being independent and identically distributed (iid) random variables uniformly sampled from $[-W/2,\, W/2]$. Indices are defined as $i = x+(y-1)L+(z-1)L^2$, where $(x,y,z)$ are Cartesian coordinates of sites and $L=V^{1/3}$ is the linear dimension. We note that the number of lattice sites is the same as the dimension of the single-particle sector of the Hilbert space. We implement open boundary conditions to avoid massive degeneracies when $W\rightarrow 0$. 
%Comparison with periodic boundary conditions for large disorder strengths is presented in App.~\ref{sec:pbc}.

At weak disorder, the model exhibits a crossover between localization in quasimomentum space (i.e., plane waves at $W=0$) and single-particle quantum chaos.
In finite systems, the crossover emerges at nonzero disorder $W_1^*$, while in the thermodynamic limit $V \rightarrow \infty$, one expects $W_1^* \rightarrow 0$.
The system is single-particle chaotic at $W>W_1^*$ in the sense that it satisfies the single-particle ETH from Eq.~\eqref{eq:eth}, and it exhibits the Wigner-Dyson distribution of single-particle level spacings~\cite{PhysRevB.47.11487, Mirlin_2000,RevModPhys.80.1355, Suntajs_2021, PhysRevB.103.104206}. 
The Anderson localization transition occurs at the critical disorder strength $W_2^*\approx 16.5$~\cite{Slevin_2018}, and it has been extensively studied in the literature~\cite{Schubert,Slevin_2014,Ohtsuki_1997,PhysRevLett.105.090601, Suntajs_2021}.
At $W>W_2^*$, the system is in a localized phase.

We consider three perturbations to the Anderson model: the kinetic energy,
\begin{equation}
\label{eq_kinetic}
    \hat{T}=-\sum_{\langle i,j \rangle} \hat{c}_{i}^\dagger \hat{c}_{j}\,,
\end{equation}
which is part of the Hamiltonian in Eq.~\eqref{eq_HA}, the sublattice kinetic energy,
\begin{equation} \label{def_Ts}
    \hat{T}_s=-\sum_{\alpha=1}^{2}\sum_{\langle\langle i,j \rangle\rangle_\alpha}\frac{1}{\alpha}\hat{c}_{i}^\dagger \hat{c}_{j}\,,
\end{equation}
where $\langle\langle i,j \rangle\rangle_{\alpha}$, with $\alpha\in\{1,2\}$, represent the sets of next-nearest neighbors, see Fig.~\ref{fig:fig0a} of Appendix~\ref{app:sublattice}, and the randomized site occupation,
\begin{equation}
\label{eq_randomized}
    \hat{n}=\sum_{i=1}^{V} (r_{i}\hat{c}_{i}^\dagger \hat{c}_{i}-\frac{r_i}{V})\,,
\end{equation}
where $r_i$ are the iid random numbers (independent from the disorder potential) drawn from the interval $[0,1]$. We generate a single configuration of $r_i$ for the largest available system size and use it in all numerical calculations.

We highlight that, although the operators from Eqs.~\eqref{eq_kinetic}–\eqref{eq_randomized} are extensive, their Hilbert--Schmidt norms in the single-particle sector of the Hilbert space, defined as $\|\hat{O}\|_{\mathrm{sp}}^{2}=\frac{1}{V}\sum_{n=1}^{V}\langle n|\hat{O}^{2}|n\rangle$, are intensive. More details concerning this choice of normalization are provided in Sec.~IID of Ref.~\cite{PhysRevB.110.104202}.

Before proceeding, we note that it is convenient to work with the rescaled fidelity susceptibilities, defined as
\begin{equation} \label{def_chi_rescaled}
\tilde{\chi}_\text{typ} = \chi_\text{typ}  \omega_\text{typ}^2 V \,,\;\;\;\; \tilde{\chi}_\text{typ/av}^r = \chi_\text{typ/av}^r \mu \;.
\end{equation}
Consequently, $\tilde{\chi}_\text{typ} \sim |f(\omega_\text{typ})|^2$ and $\tilde{\chi}_\text{typ/av}^r \sim |f(\mu)|^2$.
Since $\omega_\text{typ} \sim V^{-1}$, and assuming a similar scaling for $\mu$, these rescaled quantities are expected to decay with system size in the localized regimes, remain independent of system size when single-particle ETH is valid, and increase as
$V^a$, with $0<a\le 1$, in the vicinity of the critical points $W_1^\ast$ and $W_2^\ast$. In this section, we focus exclusively on the rescaled quantities and therefore refer to $\tilde{\chi}_\text{typ}$ and $\tilde{\chi}_\text{typ/av}^r$ from Eq.~\eqref{def_chi_rescaled} simply as “fidelity susceptibilities”. In contrast, we invoke the unscaled quantities to study the localized phase in Sec.~\ref{sec:av_versus_typ}.
We parametrize $\tilde{\chi}_\text{typ/av}^r \propto \mu^{-a}$, where $a=1-\bar a$ in Eq.~\eqref{def_bar_a}.

%%%FIGURE
\begin{figure*}[!t]
\includegraphics[width=\textwidth]{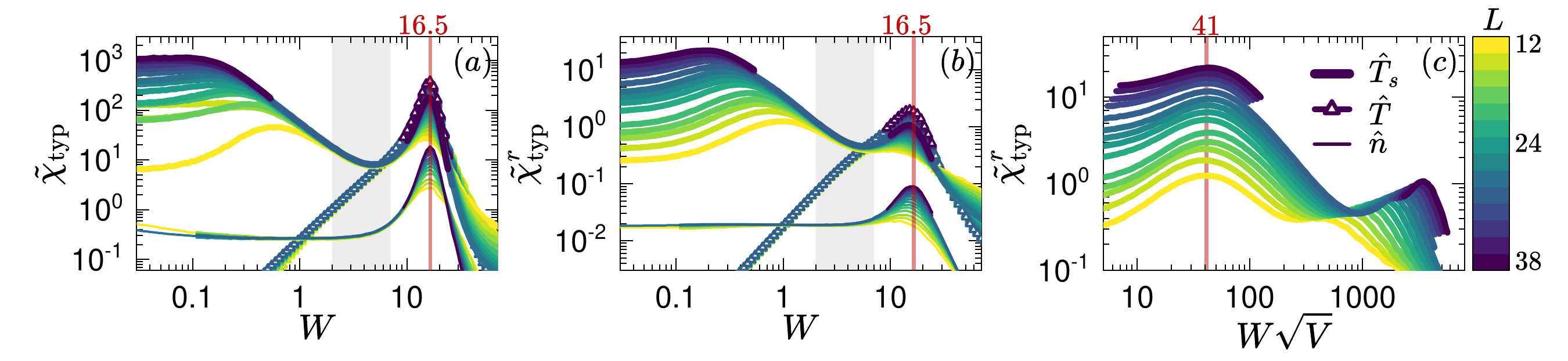}
\vspace{-0.2cm}
\caption{(a)~$\tilde{\chi}_\text{typ}$ and (b)~$\tilde{\chi}_\text{typ}^r$ as functions of $W$. (c) shows the same results as (b), but plotted vs $W\sqrt{V}$. We consider system sizes $V\le 38^3$. All numerical results were calculated from $20\%$ of single-particle energy eigenstates in the middle of the spectrum and averaged over $20$ ($5$) Hamiltonian realizations for $V\le 28^3$ ($V>28^3$). Red vertical lines in (a)-(c) mark $W_2^*=16.5$ and $W_1^*=41/\sqrt{V}$, respectively. Shaded area indicates the single-particle chaotic regime.
}
\label{fig:fig1}
\end{figure*}
%%%FIGURE

%\subsection{Scaling with system size}

We plot the unregularized typical fidelity susceptibility, $\tilde{\chi}_\text{typ}$, versus the disorder strength $W$ in Fig.~\ref{fig:fig1}(a). Its regularized counterpart, $\tilde{\chi}_\text{typ}^r$, is shown versus $W$ and $W\sqrt{V}$ in Figs.~\ref{fig:fig1}(b) and~\ref{fig:fig1}(c), respectively. Unless stated otherwise, we set the frequency cutoff $\mu=\mu^* = 2\ln(V)\omega_\text{av}$, where $\omega_\text{av}$ denotes the mean level spacing. We consider system sizes $L\le 38$, corresponding to $V\le 38^3$. All results were calculated from $20\%$ of single-particle energy eigenstates in the middle of the spectrum, and averaged over $20$ Hamiltonian realizations for $L\le 28$ and five realizations for $L>28$. The red vertical lines in Figs.~\ref{fig:fig1}(a) and~\ref{fig:fig1}(b) mark the expected critical disorder strength $W_2^*\approx 16.5$~\cite{Slevin_2018}, while the red vertical line in Fig.~\ref{fig:fig1}(c) marks $W_1^*\approx 41/\sqrt{V}$ established from second-order polynomial fits to numerical results. In the moderate disorder regime, $2\lesssim W \lesssim 7$, the fidelity susceptibilities for all operators exhibit the ETH-like scaling. In this regime, which is indicated by the shaded area in Figs.~\ref{fig:fig1}(a) and~\ref{fig:fig1}(b), single-particle chaos is expected.

It is worth commenting on why it is more convenient to scale $\mu$ with $\omega_{\text{av}}$ rather than with the previously introduced $\omega_{\text{typ}}$. For $W > W_1^*$, both level spacings behave similarly and are inversely proportional to the system size, so using $\omega_{\text{av}}$ offers no particular advantage. However, when $W \rightarrow W_1^*$, the energy spectrum begins to organize into bands of nearly degenerate states. In this regime, $\omega_{\text{typ}}$ becomes significantly smaller than $\omega_{\text{av}}$ and no longer scales inversely with the system size. This behavior is illustrated in Fig.~\ref{fig:fig2a} from Appendix~\ref{app:spectral}. 

\subsection{Weak-disorder crossover}

Interestingly, only the fidelity susceptibilities corresponding to the sublattice kinetic energy, $\hat{T}_{s}$, exhibit peaks at the weak-disorder crossover. This behavior arises because $\hat{T}_{s}$ has significant projections onto quasimomentum occupations, and hence it can be considered as an integrability-preserving perturbation~\cite{PhysRevX.10.041017,Kim_2024}. Simultaneously, the randomized site occupation, $\hat n$, has vanishing overlaps with the quasimomentum occupations, such that its off-diagonal matrix elements are roughly independent of the energy difference. Consequently, Eq.~\eqref{eq:chi_n} predicts $\chi_n \sim 1/\omega_\text{typ}$ (and $\chi_n^r \sim 1/\mu$), leading to the same scaling as in the ETH regime. Therefore, $\hat{n}$ behaves similarly to perturbations that strongly break integrability~\cite{PhysRevB.107.184312,Pozsgay_2024}. It is, nevertheless, remarkable that the corresponding fidelity susceptibility remains insensitive to the disorder strength up to very large $W\lesssim 7$. Finally, the kinetic energy, $\hat{T}$, is special because it corresponds to the total energy at $W=0$. For $n\neq m$, its offdiagonal matrix elements are
\begin{equation}
\langle n|\hat T|m\rangle=\langle n| \hat H-\sum_j \epsilon_j \hat c_j^\dagger \hat c_j|m\rangle=-\langle n| \sum_j \varepsilon_j \hat c_j^\dagger \hat c_j|m\rangle.
\end{equation}
From this expression we see that the fidelity susceptibility for $\hat T$ is similar to that for $\hat n$, except that it is multiplied by $W^2$ and therefore it vanishes as $W \to 0$.

We recall that for $W \lesssim W_1^*$, single-particle energy eigenstates become nearly translationally invariant, leading to the formation of bands of nearly degenerate single-particle energies. To mitigate this, a weak boundary disorder is introduced when performing calculations involving $\hat{T}_{s}$. More details can be found in Appendix~\ref{app:boundary}. As it is clear from Fig.~\ref{fig:fig1}, $\tilde{\chi}_\text{typ}^{r}$ shows a sharper maximum than $\tilde{\chi}_\text{typ}$ in the limit $W\rightarrow 0$. Therefore, we exclude the latter from the discussion of the weak-disorder crossover.

%%%FIGURE
\begin{figure}[!t]
\includegraphics[width=\columnwidth]{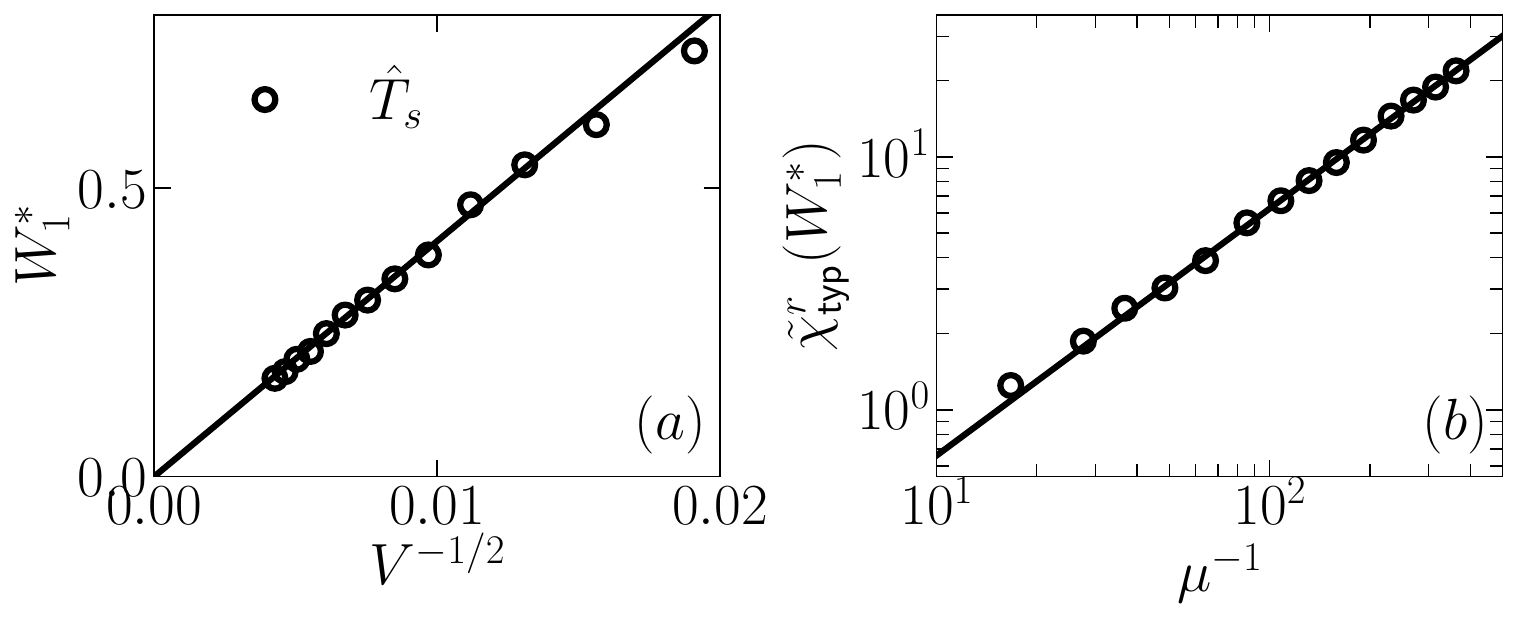}
\vspace{-0.2cm}
\caption{The weak-disorder crossover. (a)~The finite-size scaling of $W_1^*$, at which $\tilde{\chi}_\text{typ}^{r}$ is peaked, for the observable $\hat{T}_{s}$. (b)~$\tilde{\chi}_\text{typ}^{r}$ evaluated at $W_1^*$ and plotted against $\mu^{-1}$. All results were established from second-order polynomial fits to numerical results from Fig.~\ref{fig:fig1}(b). Lines in (a) and (b) are least-squares fits of $cx$ and $bx^a$ to $V \geq 18^3$, respectively. We obtain $c\approx 41$ and $a\approx0.98$.
}
\label{fig:fig2}
\end{figure}
%%%FIGURE

%%%FIGURE
\begin{figure}[!t]
\includegraphics[width=\columnwidth]{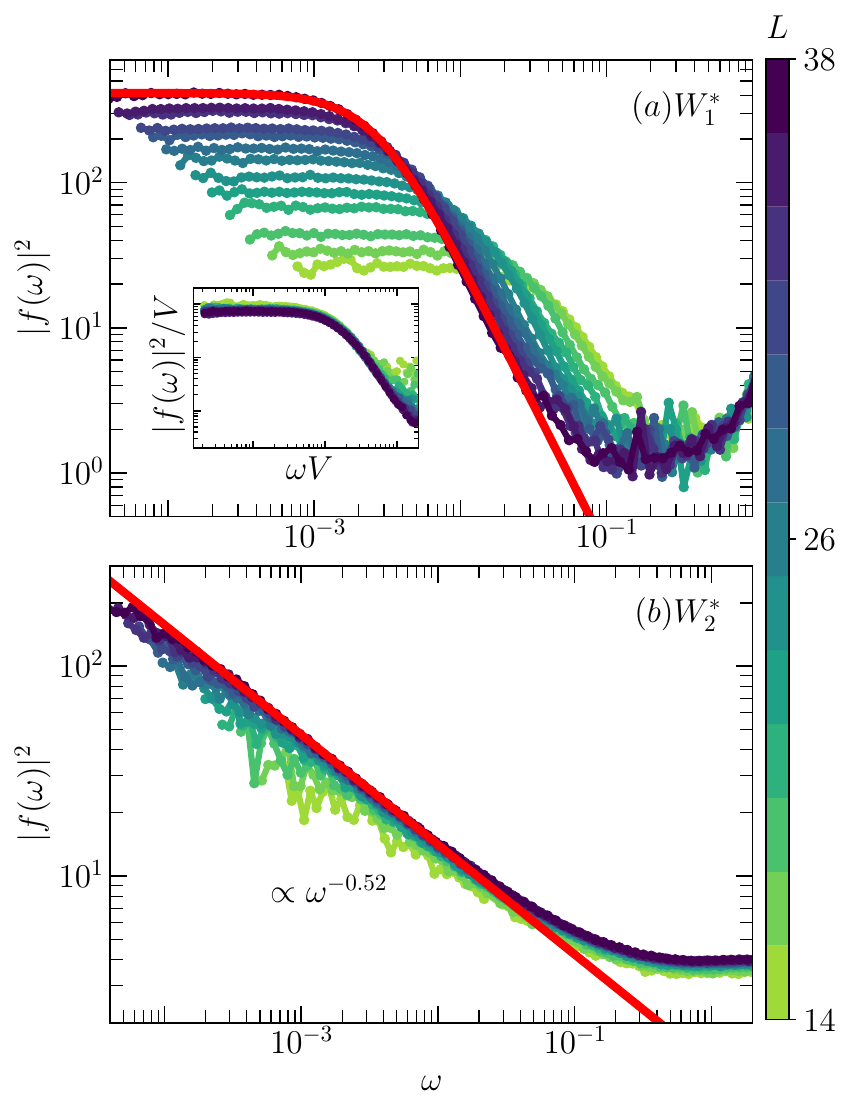}
\vspace{-0.2cm}
\caption{The spectral functions for the 3D Anderson model near (a) the weak-disorder crossover and (b) the Anderson localization transition. 
Results are obtained from $20\%$ of eigenstates in the middle of the spectrum and averaged over five disorder realizations.
The inset in (a) demonstrates the recovery of system-size independence when plotting $|f(\omega)|^2/V$ as a function of $\omega V$.
Red curves in panels (a) and (b) show the least-squares fits to the data with $V = 38^3$, using the forms
$C/(\omega^2+\Gamma^2)$ and $b\,\omega^{-a}$, respectively.
We obtain $\Gamma \approx 0.0027$ and $a \approx 0.52$.
}
\label{fig:fig3}
\end{figure}
%%%FIGURE

As expected, the peak of $\tilde{\chi}_\text{typ}^r$ for $\hat{T}_{s}$, which marks the weak-disorder crossover, shifts towards smaller $W$ as the system size increases. The peak position appears to scale with $1/\sqrt{V}$, as illustrated in Fig.~\ref{fig:fig1}(c). This is further confirmed by fitting their topmost parts with a second-order polynomial, $a_0 x^2+b_0 x+c_0$. The extracted $W_1^*=-{b_0}/{(2a_0)}$, which is plotted versus $1/\sqrt{V}$ in Fig.~\ref{fig:fig2}(a), closely follows $W_1^*\approx 41/\sqrt{V}$. We note that the disorder strength at which the level statistics begin to deviate from the Gaussian orthogonal ensemble prediction exhibit a similar drift, see Fig.~\ref{fig:fig1a} in Appendix~\ref{app:spectral}.

The observed behavior is consistent with the scaling of the relaxation rate in nearly-integrable regimes, where $\Gamma$ follows the Fermi golden rule and is proportional to the square of integrability-breaking perturbation~\cite{PhysRevB.76.245108, PhysRevB.92.195121, PhysRevLett.115.180601, PhysRevLett.120.070603, Mierzejewski_2022}. 
In the 3D Anderson model, this gives $\Gamma \sim W^2$. Consequently, the weak-disorder crossover emerges when $\Gamma$ becomes comparable to $\omega_\text{H} = \omega_\text{typ}\sim V^{-1}$, leading to $W_1^* \sim 1/\sqrt{V}$. 

This discussion suggests that the peak in $\tilde{\chi}_\text{typ}^r$ can be interpreted in terms of fading ergodicity, and hence it corresponds to scenario 1 discussed in Sec.~\ref{sec:peak_scaling}.
This can be demonstrated by calculating the spectral function,
\begin{equation}
\label{eq:spectral}
    |f(\omega)|^2 \approx \frac{Z}{||\Lambda(\omega)||} \sum_{\substack{\ket{n},\ket{m}\in\Lambda \\ \omega_{nm}\in[\omega-\delta\omega,\omega+\delta\omega)}} |\langle n | \hat{T}_s | m \rangle|^2,
\end{equation}
where $\Lambda$ is a set comprising $20\%$ of single-particle energy eigenstates from the middle of the spectrum, $\delta\omega$ is established after dividing the range of $\omega$ into $100$ bins on a logarithmic scale, and $||\Lambda(\omega)||$ is the number of off-diagonal matrix elements in the bin around $\omega$. We then average $|f(\omega)|^2$ over $20$ disorder realizations. 

We show $|f(\omega)|^2$ in Fig.~\ref{fig:fig3}(a) for $V \le 38^3$. Each curve corresponds to a numerically determined $W_1^*$, at which $\tilde{\chi}_{\text{typ}}^r$ exhibits a peak for a given $V$.
The spectral functions are clearly Lorentzian, as demonstrated by the least-squares fit to $C/ (\omega^2 + \Gamma^2)$ to the data for $V = 38^3$, see the red curve in Fig.~\ref{fig:fig3}(a). The fit yields $\Gamma \approx 0.0027$. Moreover, the spectral functions change with the system size, as the height of the plateau increases with $V$. This is further confirmed in the inset of Fig.~\ref{fig:fig3}(a), where the system-size independence is recovered when $|f(\omega)|^2/V$ is plotted against $\omega V$. 
These scaling properties are analogous to those observed in systems that comply with fading ergodicity~\cite{Kliczkowski_2024}.
Consequently, the fluctuations of the off-diagonal matrix elements at low $\omega$ increase with $V$ at the critical $W_1^*$, giving rise to the enhancement of fidelity susceptibility.
In Fig.~\ref{fig:fig2}(b), we show that $\tilde{\chi}_\text{typ}^r$ at $W_1^*$ scales as $\propto\mu^{-a}$, with $a = 0.98$, and hence it is close to being maximally divergent. 
Fading ergodicity predicts that the fidelity susceptibility is maximal when $\Gamma=\mu$, so that $\chi_\text{typ}^r\sim\frac{1}{8\mu}$, see Eq.~\eqref{eq:chi_Gamma}, which is consistent with the results in Fig.~\ref{fig:fig2}(b).

We note that when studying $\tilde{\chi}^r_\text{typ}$ with $\mu \neq 0$, two distinct scales are relevant: the frequency cutoff $\mu$, and the system size $V$ (which can emerge as an independent scale rather than through the definition of $\mu$). The system size $V$ naturally enters the picture when the frequency cutoff $\mu$ is close to the typical level spacing $\omega_\text{typ}$, as $\omega_\text{typ} \sim V^{-1}$. Nevertheless, the system size can also enter through a faster ballistic scale $v/L$, where $v$ is the Lieb-Robinson velocity~\cite{LiebRobinson}. We are generally interested in the regime where $\omega_{\rm typ}\ll \mu\ll v/L$. The interplay between these two scales, which is revealed in different behaviors of $\tilde{\chi}^r_\text{typ}$ for large versus small $\mu V$, is discussed in detail in Sec.~\ref{sec:scaling_mu}.

\subsection{Anderson localization transition} \label{sec:anderson_transition}

We now shift our focus to the Anderson localization transition, at which fidelity susceptibilities develop peaks for all considered perturbations.
Results in Fig.~\ref{fig:fig1}(a) and~\ref{fig:fig1}(b) show that the position of the peaks of $\tilde{\chi}_\text{typ}$ and $\tilde{\chi}_\text{typ}^r$, respectively, indeed emerges very close to the predicted critical point $W_2^*=16.5$, with only weak dependence on the system size. 
This is the first main result of this section.

\begin{figure}[!t]
\includegraphics[width=\columnwidth]{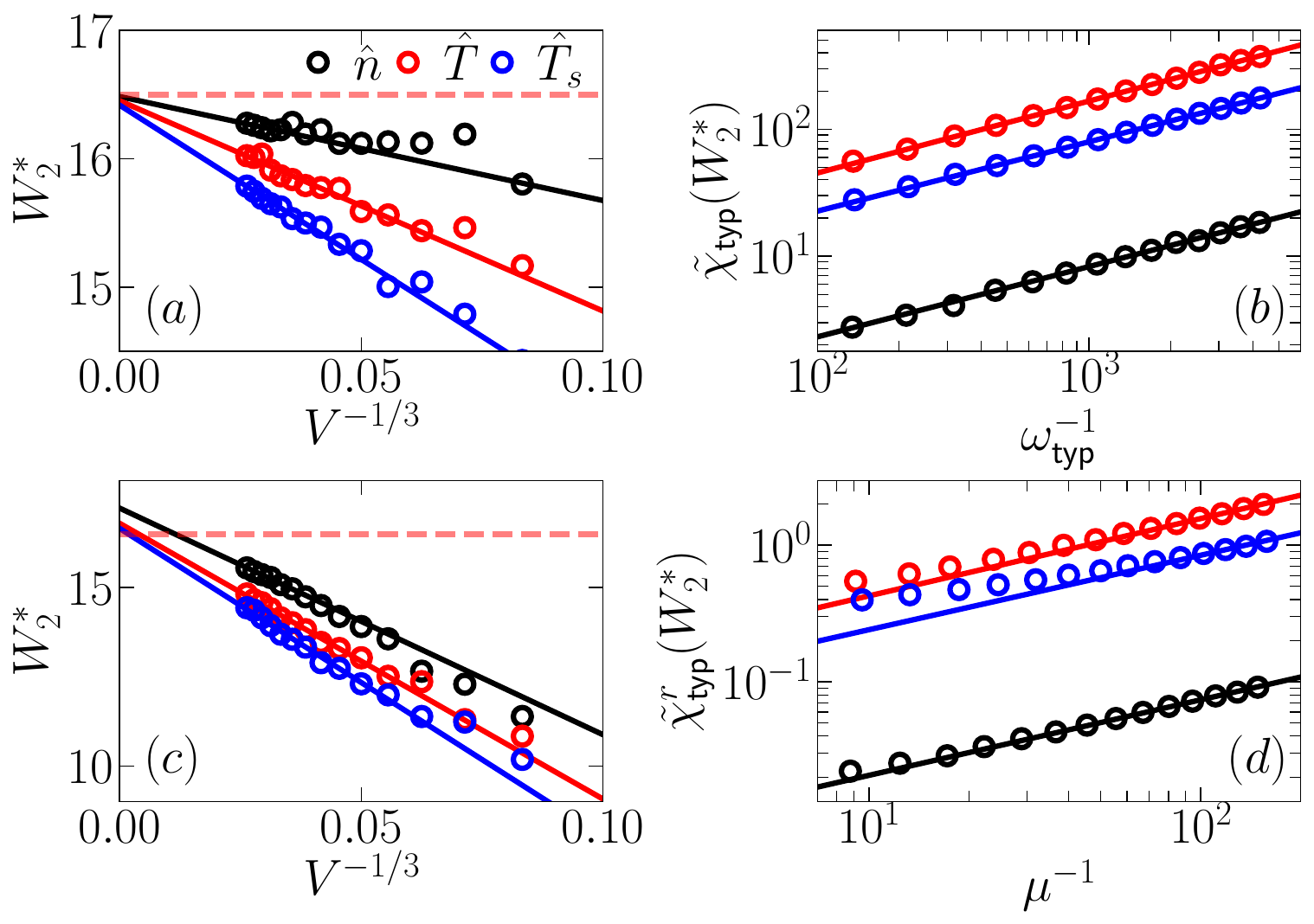}
\vspace{-0.2cm}
\caption{The Anderson localization transition. Results in (a), (b) and (c), (d) were established from second-order polynomial fits to numerical results from Figs.~\ref{fig:fig1}(a) and~\ref{fig:fig1}(b), respectively. The finite-size scaling of $W_2^*$, at which (a) $\tilde{\chi}_\text{typ}$ and (c) $\tilde{\chi}_\text{typ}^{r}$  are peaked. (b)~$\tilde{\chi}_\text{typ}$ evaluated at $W_2^*$ and plotted versus $\omega_\text{typ}^{-1}$. (d)~$\tilde{\chi}_\text{typ}^{r}$ evaluated at $W_2^*$ and plotted versus $\mu^{-1}$. Lines in panels [(a) and (c)] (b) are least-squares fits of [$dx + c$] $bx^a$, respectively. These fits are performed for $V \geq 18^3$ [with the exception of $\hat{n}$ in panels (a) and (b), where $V \geq 30^3$ is considered]. For $\hat{T}_{s},\hat{T}$ and $\hat{n}$, we obtain $c=16.42,16.45,16.48$ in (a), $c=16.66,16.76,17.23$ in (c) and $a=0.546,0.567,0.554$ in (b). The uncertainties are comparable to or smaller than the symbol size for all $L\geq 14$. Their average relative values in (a)-(d) are below $0.5\%$, $2\%$, $2\%$ and $1\%$, respectively.
Lines in (d) are guides to the eye and follow power laws with the same exponents as those established in (b). Horizontal lines in (a) and (c) mark $W_2^*=16.5$.
}
\label{fig:fig4}
\end{figure}

We fit the topmost parts of the peaks of $\tilde{\chi}_\text{typ}$ and $\tilde{\chi}_\text{typ}^r$ using a second-order polynomial, $a_0 x^2+b_0 x+c_0$, and we plot the finite-size scaling of $W_2^* = -b_0/(2a_0)$ in Figs.~\ref{fig:fig4}(a) and~\ref{fig:fig4}(c), respectively. The uncertainties of $W_2^*$ have been calculated as $u(W_2^*)^2=\sum_{i}[(\partial_{i}W_2^*)^2\text{Var}(i)+\sum_{j\neq i}\partial_{i}W_2^*\partial_{j}W_2^*\text{Cov}(i,j)]$, where $i,j\in\{a_0,b_0,c_0\}$, while $\text{Var}(i)$ and $\text{Cov}(i,j)$ are the diagonal and off-diagonal elements of the covariance matrix from the fit, respectively. The uncertainties are comparable to or smaller than the symbol size for all $L\geq 14$. Their average relative values in panels (a) and (c) are below $0.5\%$ and $2\%$, respectively.
For $\tilde{\chi}_\text{typ}$, the critical disorder strength closely follows $W_2^*\approx c + d V^{-1/3}$, which enables an extrapolation to the infinite system-size limit. We obtain $W_2^* \rightarrow 16.42,16.45,16.48$ for $\hat{T}_s$, $\hat{T}$, and $\hat{n}$, see Fig.~\ref{fig:fig4}(a). 
The agreement with the expected $W_2^* \approx 16.5$ is striking, with a relative error below $1\%$. 
The same scaling ansatz also appears to apply for $\tilde{\chi}_\text{typ}^r$.
We obtain $W_2^* \rightarrow 16.66$, $16.76$, and $17.23$ for $\hat{T}_s$, $\hat{T}$, and $\hat{n}$, respectively, see Fig.~\ref{fig:fig4}(c). Interestingly, all offsets are greater than the expected $W_2^*$, and their relative errors are larger than for $\tilde{\chi}_\text{typ}$, reaching up to $4.5\%$ for $\hat{n}$. 

We assume that the small quantitative differences between the positions of the peaks of $\tilde{\chi}_\text{typ}$ and $\tilde{\chi}_\text{typ}^r$ are finite-size effects. 
It appears that the fidelity susceptibility, and hence $W_2^\ast$, depends both on $\mu$ and the dimensionless argument $\mu V$, which will be further discussed in Sec.~\ref{sec:scaling_mu}.
This two-parameter scaling of fidelity susceptibilities explains a slight variation of the slopes of the observables, especially of $\hat n$, visible in Fig.~\ref{fig:fig4}(c).
Nevertheless, we expect the critical disorder strength extracted from $\tilde{\chi}_\text{typ}^r$ to reproduce the correct $W_2^*$ in the thermodynamic limit.

%%%FIGURE
%%%FIGURE

The second main result is that the height of the peak of fidelity susceptibilities at the Anderson localization transition is significantly more slowly than the maximal rate, as shown in Figs.~\ref{fig:fig4}(b) and~\ref{fig:fig4}(d). We emphasize that the uncertainties of $\tilde{\chi}_\text{typ}$ have been calculated in the same manner as those of $W_2^*$. The uncertainties are comparable to or smaller than the symbol size for all $L\geq 14$. Their average relative values in panels (b) and (d) are below $2\%$ and $1\%$, respectively. In particular, $\tilde{\chi}_\text{typ}$ evaluated at $W_2^*$ increases as $\propto\omega_\text{typ}^{-a}$, with $a = 0.546,0.567$ and $0.554$ for $\hat{T}_s$, $\hat{T}$, and $\hat{n}$, respectively, see Fig.~\ref{fig:fig4}(b). 
The origin of such scaling will be discussed below. At the same time, the regularized $\tilde{\chi}_\text{typ}^r$ appears to exhibit a correction to a simple power law, which becomes less significant for smaller $\mu\sim \ln V / V$. 
This is another indication of a two-parameter scaling. In the limit of vanishing $\mu$ (but diverging $V$ and $\mu V$), they seem to follow the same scaling behavior as observed for the unregularized $\tilde{\chi}_\text{typ}$. This is illustrated in Fig.~\ref{fig:fig4}(d), where the lines are guides to the eye and represent $b\mu^{-a}$, with the exponents $a$ matching those extracted from Fig.~\ref{fig:fig4}(b). 

We complement these results by analyzing the spectral functions, $|f(\omega)|^2$ from Eq.~\eqref{eq:spectral}, associated with the sublattice kinetic energy, $\hat{T}_s$. Results are shown in Fig.~\ref{fig:fig3}(b) for $V \le 38^3$. The curve for each system corresponds to a numerically extracted $W_2^*$, at which $\tilde{\chi}_\text{typ}$ develops a peak.
We observe that $|f(\omega)|^2$ are scale-invariant and develop a polynomially decaying tail extending over a broad range of $\omega$. This tail closely follows the function $\propto\omega^{-a}$, with $a \approx 0.52$, see the red curve in Fig.~\ref{fig:fig3}(b).
(Note that we obtained $a\approx0.546$ from the behavior of $\tilde{\chi}_\text{typ}$ for $\hat{T}_s$, which is within $5\%$ of the relative error. Moreover, the exponent of the tail varies slightly with the fitting range.) We highlight that a similar power-law behavior of $|f(\omega)|^2$ has been observed in the classical limit of the two-spin XYZ model near the integrability-breaking transition~\cite{Lim_2024}. In Fig.~\ref{fig:fig3a} of Appendix~\ref{app:spectral}, we also investigate, for all considered perturbations, how $|f(\omega)|^2$ evolves as the disorder strength varies across the Anderson localization transition.

We argue that the sub-maximal scaling of the peaks of fidelity susceptibilities, i.e., $a<1$, is related to the power-law decay of the autocorrelation function of site occupation, $\hat{n}_{i}=\sqrt{V}\hat{c}^\dagger_i\hat{c}_i-1/\sqrt{V}$, and the survival probability, $|\langle i | i(t) \rangle|^2$~\footnote{When defining $\hat{n}_{i}$, we ensured that it is traceless and normalized within the single-particle sector of the Hilbert space~\cite{PhysRevB.104.214203}.}. In the single-particle systems they are connected via $1/V\sum_{n}\langle n | \hat{n}_{i}\hat{n}_{i}(t)|n\rangle_c = |\langle i | i(t) \rangle|^2 + {\rm const}$. 
It is known that at the localization transition of the 3D Anderson model, the survival probability decays as $\propto t^{-d_2}$, where $d_2 \approx 0.42$ is the wavefunction fractal dimension in the site-occupation basis~\cite{Chalker88, Ketzmerick_92, Huckestein94, Ketzmerick_97, cuevas_kravtsov_07, Kravtsov10, PhysRevLett.131.060404, Hopjan_2023, jiricek_hopjan_26b}, which can be calculated via the scaling properties of the inverse participation ratio~\cite{PhysRevLett.131.060404,PhysRevB.84.134209,PhysRevB.91.184206}. We note that $0<d_2<1$, with the smallest (largest) value attained for localized (delocalized) states. In the literature, a related exponent is often introduced, $D_2=d d_2$, where $d$ is the system dimension. It can be expressed as $D_2=d-\Delta_2$, where $\Delta_2$ quantifies the deviation of $D_2$ from its maximal value, $D_2=d$, which is reached for delocalized states. We highlight the existence of very early numerical studies of $D_2$~\cite{PhysRevLett.83.4590,PhysRevLett.52.565,Brandes1996,PhysRevLett.67.607}.
%at least when $|\langle i | i(t) \rangle|^2$ is averaged over $i$
This results in $|f(\omega)|^2 \sim \omega^{-(1-d_2)}$, and hence it predicts $\tilde{\chi}_\text{typ}\sim \omega_\text{typ}^{-(1-d_2)}$ and $\tilde{\chi}_\text{typ}^r\sim \mu^{-(1-d_2)}$, i.e., $a=1-d_2$. The latter exponent can also be expressed as $a=\Delta_2/d$.

We note that the argument above was carried out for $\hat{n}_i$ and, consequently, it predicts the behavior of spectral functions and fidelity susceptibilities for $\hat{n}$. Nevertheless, we expect that all considered perturbations couple similarly to a large number of slow modes, resulting in their autocorrelation functions exhibiting similar time evolutions, see Appendix~\ref{app:sublattice} and also Ref.~\cite{PhysRevLett.128.190601}. The predicted scalings are consistent with the findings of this manuscript, e.g., we obtain $d_2\approx 1- 0.554=0.446$ from the behavior of $\tilde{\chi}_\text{typ}$ for $\hat{n}$, which corresponds to the relative discrepancy of approximately $6\%$ with the earlier predictions. 

Before moving on, we note that the correlation functions commonly studied in the context of Anderson localization transition, $S(r,\omega)$, which scaling is known to be controlled by $\Delta_2/d$~\cite{Brandes1996}, are related to the autocorrelation function of $\hat{n}_i=\hat{c}_i^\dagger\hat{c}_i$ in the limit $r\rightarrow0$. For other one-body operators, there is no straightforward relation.

\subsection{Scaling with frequency cutoff} \label{sec:scaling_mu}

\begin{figure}[!t]
\includegraphics[width=\columnwidth]{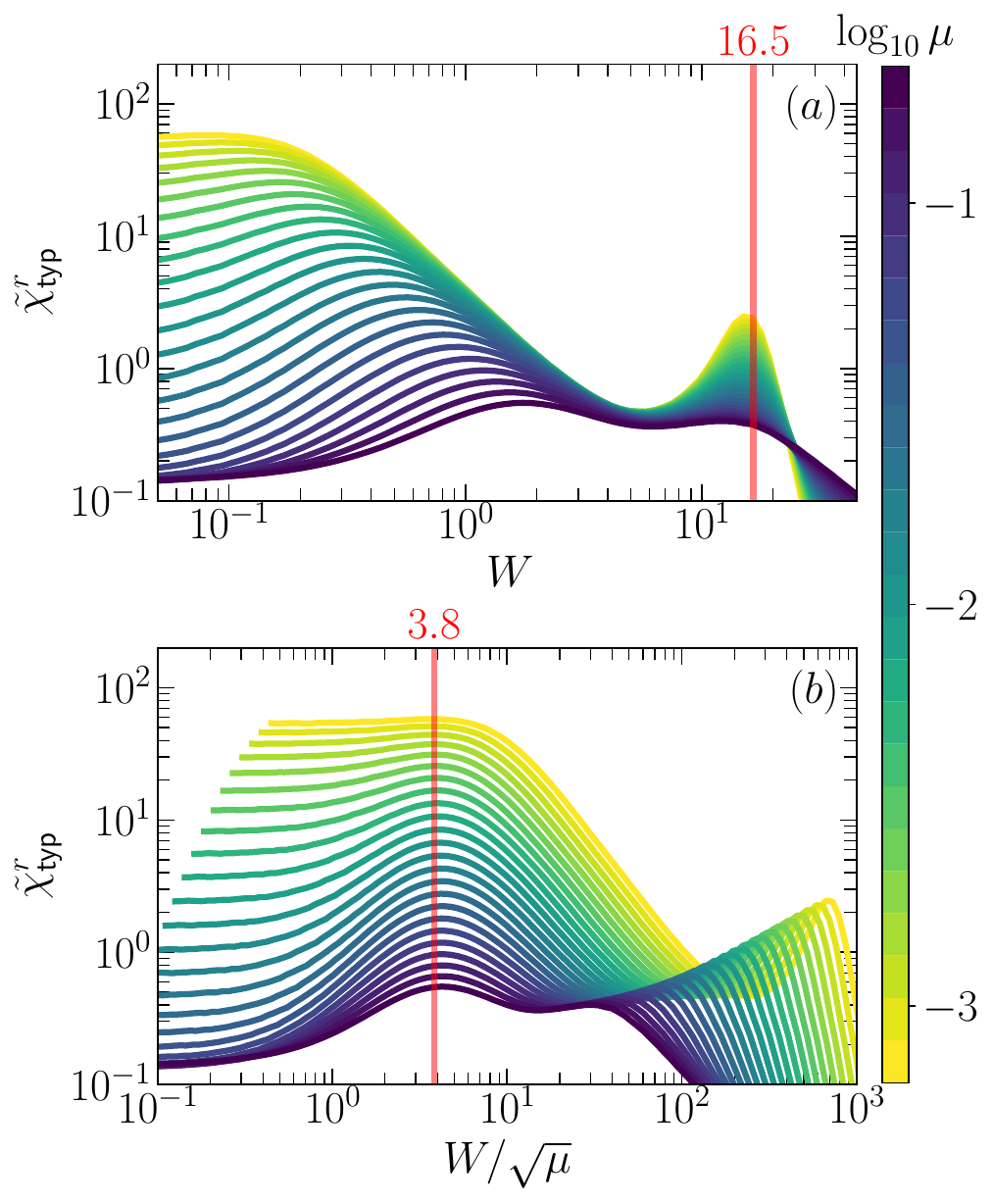}
\vspace{-0.2cm}
\caption{$\tilde{\chi}_\text{typ}^r$ of observable $\hat T_s$ as functions of (a) $W$ and (b) $W/\sqrt{\mu}$ for $V=38^3$.  We consider $10^{-4}<\mu<1$, and $\mu$ is independent of $W$. All numerical results were calculated from $20\%$ of single-particle energy eigenstates in the middle of the spectrum and averaged over $5$ Hamiltonian realizations. Red vertical lines in (a) and (b) mark $W_2^*\approx16.5$ and $W_1^* \approx 3.8 \sqrt{\mu}$, respectively.
}
\label{fig:fig5}
\end{figure}

We now comment on how the fidelity susceptibility scales with the frequency cutoff $\mu$, keeping the system size fixed. To simplify the analysis, we only consider the sublattice kinetic energy $\hat{T}_s$ as a perturbation. 
First, we plot $\tilde{\chi}_\text{typ}^r$ versus $W$ for $L=38$ in Fig.~\ref{fig:fig5}(a). We consider $10^{-4}<\mu<1$ and darker colors indicate larger $\mu$. We emphasize that, in contrast to Fig.~\ref{fig:fig1}, $\mu$ is independent of $W$. As before, all results were
calculated from $20\%$ of single-particle eigenstates in the middle of the spectrum, and averaged over five Hamiltonian realizations. The red vertical line in Fig.~\ref{fig:fig5}(a) marks the prediction for the Anderson localization transition, $W_2^*\approx16.5$. 

%%%FIGURE
%%%FIGURE

%%%FIGURE
\begin{figure}[!t]
\includegraphics[width=\columnwidth]{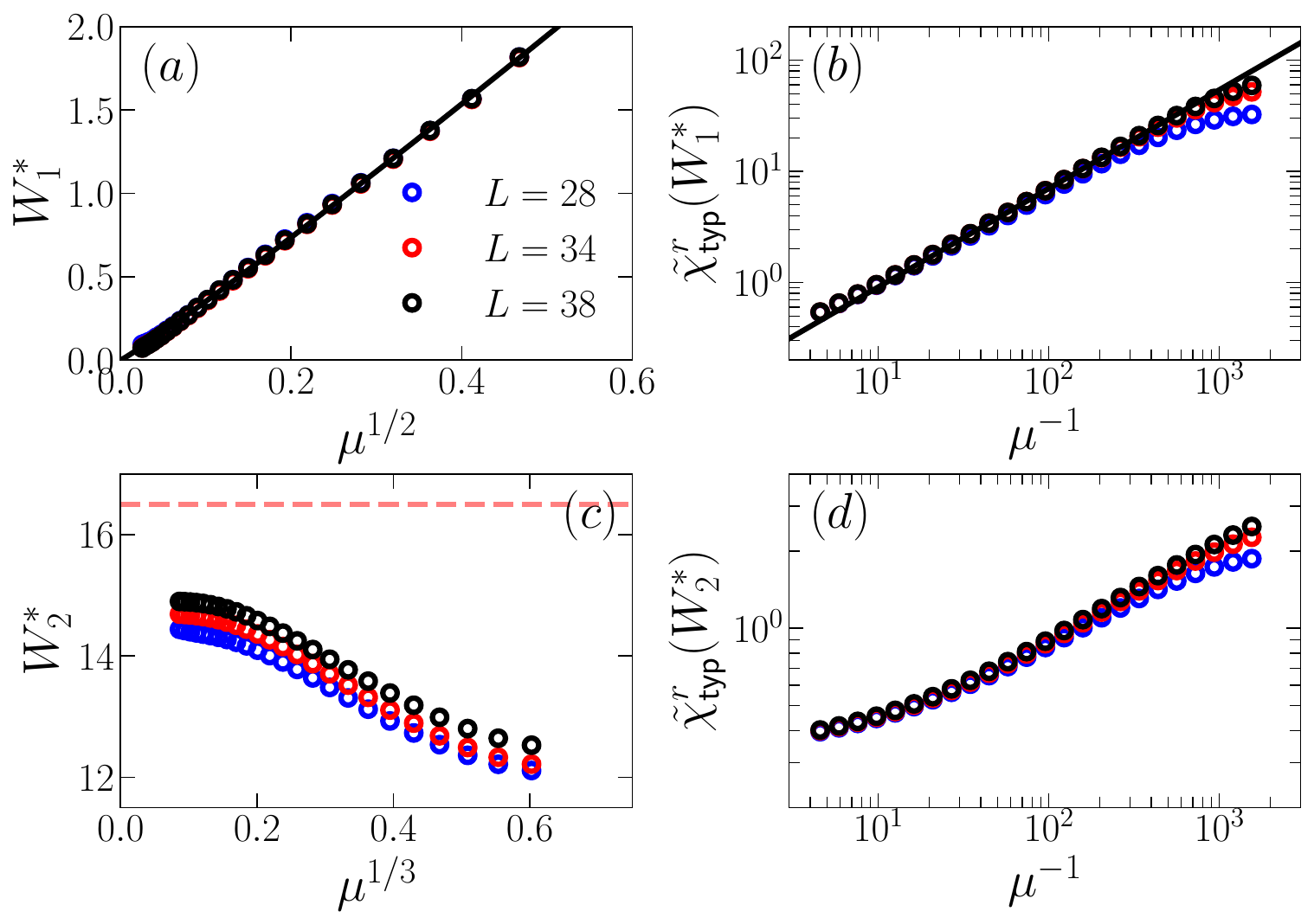}
\vspace{-0.2cm}
\caption{The weak-disorder crossover: (a) the peak position, $W_1^*$, vs $\mu^{1/2}$, and (b) the peak height, $\tilde{\chi}_\text{typ}^{r}(W_1^*)$, vs $\mu^{-1}$. The Anderson localization transition: (c) the peak position, $W_2^*$, vs $\mu^{1/3}$ and (d) the peak height, $\tilde{\chi}_\text{typ}^{r}(W_2^*)$, vs $\mu^{-1}$. We consider $V=28^3,34^3$ and $38^3$. All results were obtained from second-order polynomial fits to numerical data, analogous to those shown in Fig.~\ref{fig:fig5}. The black lines in (a) and (b) show the least-squares fits of $b x^{a}$ applied to the power-law regions (moderate $\mu$) of the plots for $V = 38^3$. The red dashed line in (c) marks the expected $W_2^*\approx 16.5$.
}
\label{fig:fig6}
\end{figure}
%%%FIGURE

%%%FIGURE
\begin{figure*}[!t]
\includegraphics[width=\textwidth]{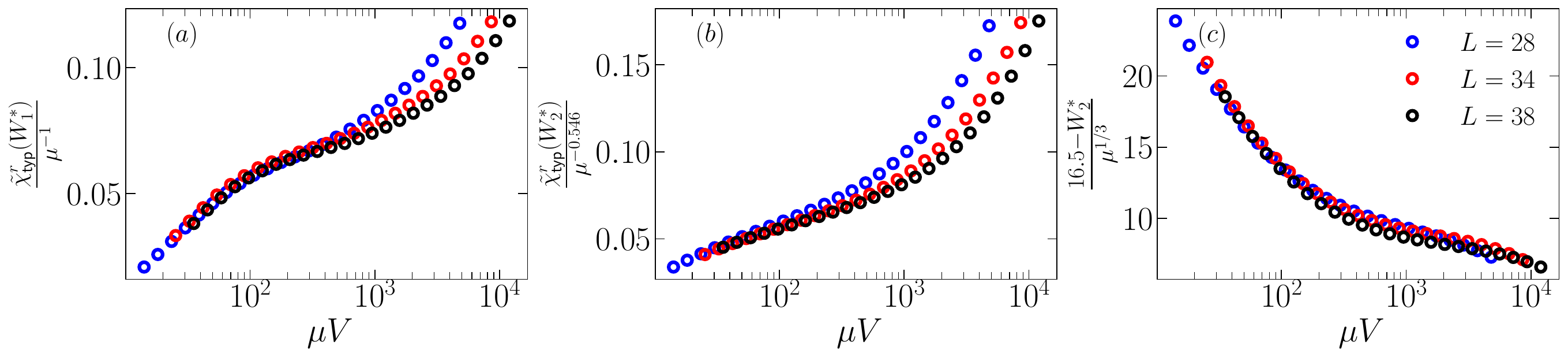}
\vspace{-0.2cm}
\caption{The weak-disorder crossover: (a) $\tilde{\chi}_{\text{typ}}^{r}(W_1^*)/\mu^{-1}$ plotted vs $\mu V$. The scaling collapse is achieved in the limit $\mu \rightarrow \omega_{\text{typ}}$. The Anderson localization transition: (b) $\tilde{\chi}_{\text{typ}}^{r}(W_2^*)/\mu^{-0.546}$ and (c) $(16.5 - W_2^*) / \mu^{1/3}$ plotted vs $\mu V$. In the latter case, the scaling collapse is observed for all considered $\mu$, whereas in the former it occurs in the limit $\mu \rightarrow \omega_{\text{typ}}$. We consider $V=28^3,34^3$ and $38^3$, an observable $\hat T_s$.}
\label{fig:fig7}
\end{figure*}
%%%FIGURE

The variation of $\tilde{\chi}_{\text{typ}}^r$ with decreasing $\mu$, see Fig.~\ref{fig:fig5}(a), resembles its variation with increasing $V$, see Fig.~\ref{fig:fig1}(b). 
This is not particularly surprising, as in the previous section the system size $V$ entered the unregularized $\tilde{\chi}_\text{typ}$ primarily through the Heisenberg frequency $\omega_{\rm typ} \sim 1/V$, while it entered the regularized $\tilde{\chi}_\text{typ}^r$ through the system-size dependence of the frequency cutoff $\mu \sim \log(V)/V$. This is further supported by Fig.~\ref{fig:fig5}(b), where $\tilde{\chi}_\text{typ}^r$ is plotted versus $W/\sqrt{\mu}$, resulting in the alignment of the first peak for different values of $\mu$. The red vertical line in Fig.~\ref{fig:fig5}(b) marks $W_1^* \approx 3.8 \sqrt{\mu}$, obtained from the second-order polynomial fits to the numerical results. Note that the variation of $\tilde{\chi}_\text{typ}^r$ with $\mu$ can be accessed experimentally by probing the nonequilibrium dynamics of a system at time $t = \mu^{-1}$, which is far shorter than the longest quantum time scale $\tau_{\text{H}} = \omega_{\text{typ}}^{-1}$.

The signatures of a two-parameter scaling of fidelity susceptibility near the Anderson localization transition were already pointed out in Sec.~\ref{sec:anderson_transition}. A closer examination of Fig.~\ref{fig:fig5} reinforces this observation. For example, the peak at $W_1^*$ in Fig.~\ref{fig:fig5}(b) begins to shift with respect to $W^*\approx 3.8\sqrt{\mu}$ at sufficiently small $\mu$ (and $\mu V$). The effect becomes apparent when we perform the second-order polynomial fits around the peaks in the fidelity susceptibility, and plot $W_1^*$ and $\tilde{\chi}_{\text{typ}}^{r}(W_1^*)$ in Figs.~\ref{fig:fig6}(a) and \ref{fig:fig6}(b), respectively. We also plot $W_2^*$ and $\tilde{\chi}_{\text{typ}}^{r}(W_2^*)$ in Figs.~\ref{fig:fig6}(c) and \ref{fig:fig6}(d), respectively. We consider three system sizes, i.e., $L=28,\;34$ and $38$. 

Focusing on the weak-disorder crossover in Figs.~\ref{fig:fig6}(a) and~\ref{fig:fig6}(b), we find that both the position and the height of the peak of $\tilde{\chi}_{\text{typ}}^r$ are determined by $\mu$, provided that $\mu V$ is not too small or too large, i.e., $ \omega_\text{typ}\sim V^{-1} \ll \mu\ll 10^{-1}$. In particular, we obtain the critical disorder strength $W_1^* \sim \sqrt{\mu}$ and the corresponding fidelity susceptibility $\tilde{\chi}_{\text{typ}}^{r}(W_1^*) \sim \mu^{-a}$ with $a \approx 0.89$.   
The deviation at large $\mu$ can be easily explained, as $|f(\omega)|^2$ does not follow a Lorentzian function in this regime of frequencies for the considered system sizes, see Fig.~\ref{fig:fig3}(a).
When $\mu V$ is small ($\mu \approx \omega_{\text{typ}} \sim V^{-1}$), $\tilde{\chi}_{\text{typ}}^{r}(W_1^*)$ acquires an additional dependence on $V$.
In Fig.~\ref{fig:fig7}(a), we show evidence that $\tilde{\chi}_{\text{typ}}^{r}(W_1^*)$ can be expressed as $\mu^{-1} g_1(\mu V)$ for small $\mu V$, where $g_1(\mu V)$ is a smooth function of $\mu V$. For moderate $\mu V$, $g_1(\mu V) \approx \text{const.}$, and, as expected, the results are consistent with an increasing range of validity of scaling collapse with increasing volume $V$.

The situation is different at the Anderson localization transition. Although the behavior of the peak height, $\tilde{\chi}_{\text{typ}}^{r}(W_2^*)$, is governed by $\mu$ for not too small $\mu V$, it no longer follows the scaling ansatz $\mu^{-a}$, see Fig.~\ref{fig:fig6}(d). This observation is consistent with the results from previous sections, where we established that the unregularized fidelity susceptibility, $\tilde{\chi}_\text{typ}$, is well approximated by $\omega_\text{typ}^{-0.546}$, see Fig.~\ref{fig:fig4}(b). In particular, $\tilde{\chi}_{\text{typ}}^{r}(W_2^*)$ can be expressed as $\mu^{-0.546} g_2(\mu V)$ for small $\mu V$, where $g_2(\mu V)$ stands for a smooth function of $\mu V$, as shown in Fig.~\ref{fig:fig7}(b).

Finally, Fig.~\ref{fig:fig6}(c) suggests that $W_2^*$ is not proportional to $\mu^{1/3}$ and, moreover, it drifts towards larger $W$ with increasing $V$. To connect to the results from the previous sections, we seek for the scaling collapse by plotting $(16.5 - W_2^*) / \mu^{1/3}$ versus $\mu V$. Interestingly, this scaling collapse works not only when $\mu$ approaches $\omega_{\text{typ}}$, but also for larger $\mu$, as demonstrated in Fig.~\ref{fig:fig7}(c).

%%%%%%%%%%%%%%%%%%%%%%%%%%%%%%%%%%%%%%%%%

%%%%%%%%%%%%%%%%%%%%%%%%%%%%%%%%%%%%%%%%%
\section{Fidelity susceptibility in \\localized phase}
\label{sec:av_versus_typ}

\subsection{Average versus typical fidelity susceptibility} \label{sec:av_typical}

\begin{figure}[!t]
\includegraphics[width=\columnwidth]{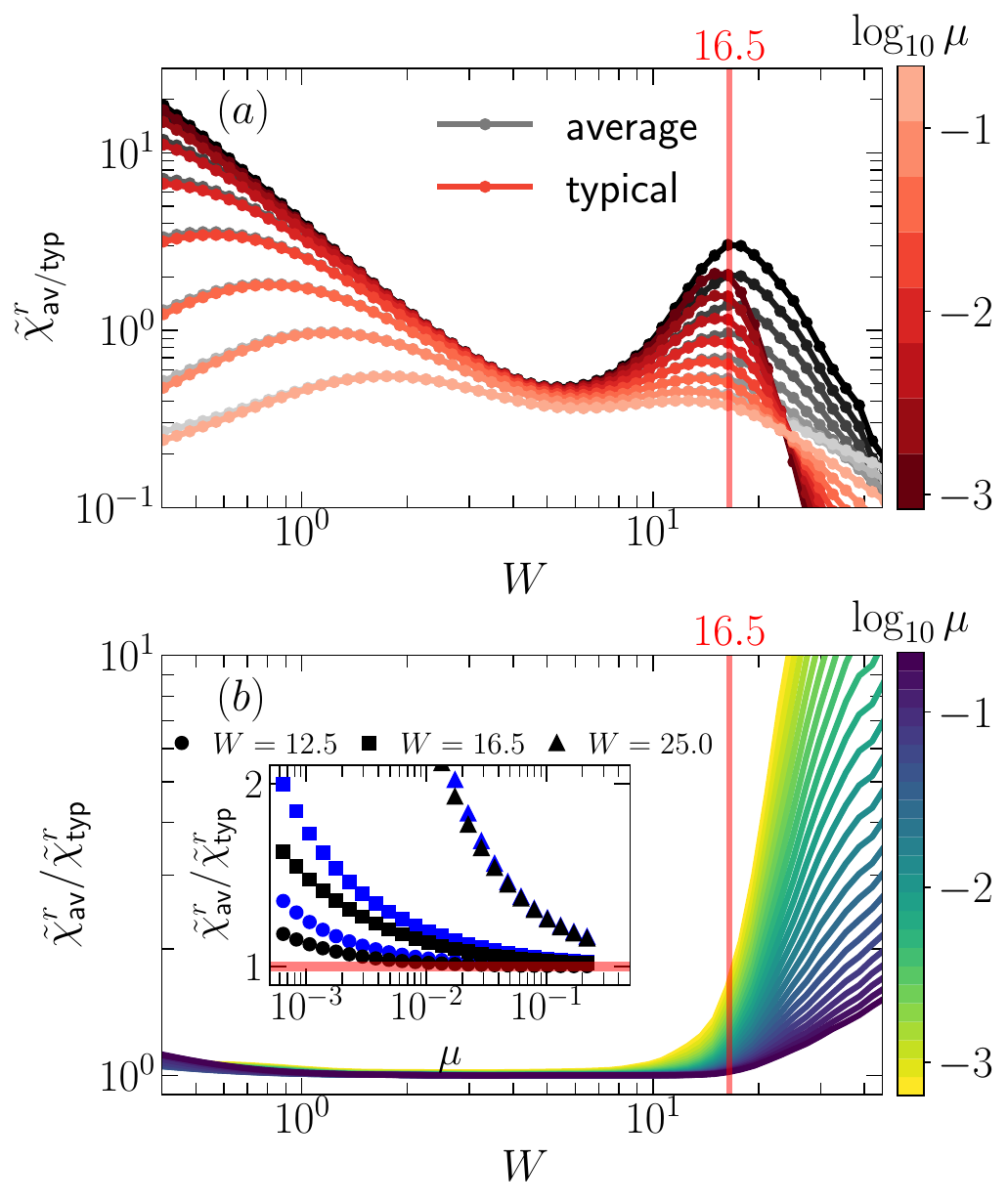}
\vspace{-0.2cm}
\caption{(a) $\tilde{\chi}_\text{av}^{r}$ (grey color scale) and $\tilde{\chi}_\text{typ}^{r}$ (red color scale) as functions of $W$, for observable $\hat T_s$. (b) $\tilde{\chi}_\text{av}^{r}/\tilde{\chi}_\text{typ}^{r}$ plotted against $W$. We consider $V=38^3$ as well as $10^{-4}<\mu<1$, where $\mu$ is independent of $W$. The inset shows $\tilde{\chi}_\text{av}^{r}/\tilde{\chi}_\text{typ}^{r}$ plotted vs $\mu$ for three disorder strengths: $W=12.5$ (circles), $16.5$ (squares), and $25.0$ (triangles). Blue symbols correspond to $V=28^3$, while black symbols to $V=38^3$. All results have been calculated from $20\%$ of energy eigenstates in the middle of the spectrum and 5 disorder realizations. Red vertical lines in (a) and (b) mark the expected $W_2^*\approx 16.5$, while the red horizontal line in the inset marks $\tilde{\chi}_\text{av}^{r}/\tilde{\chi}_\text{typ}^{r}=1$.
}
\label{fig:fig8}
\end{figure}

So far, we have only considered the typical fidelity susceptibility.
We now comment on the differences between the average and typical fidelity susceptibilities. In Fig.~\ref{fig:fig8}(a), $\tilde{\chi}_\text{av}^{r}$ and $\tilde{\chi}_\text{typ}^{r}$ of observable $\hat T_s$, are plotted against $W$ using grey and red color scales, respectively. We consider one system size, $V=38^3$, along with different frequency cutoffs, $10^{-4} < \mu < 1$, which are independent of disorder strength, $W$. Fixing the frequency cutoff, $\mu$, and varying the system size, $V$, yields qualitatively similar results (not shown).

%%%FIGURE
%%%FIGURE

It is clear from Fig.~\ref{fig:fig8}(a) that in the single-particle chaotic regime, i.e., at $W_1^\ast <W<W_2^\ast$, both quantities remain very close, almost overlapping. Below the weak-disorder crossover, $W < W_1^\ast$, the typical value $\chi_{\mathrm{typ}}^{r}$ decays slightly faster with decreasing disorder strength than the average $\chi_{\mathrm{av}}^{r}$, suggesting the existence of eigenstates with anomalously large $\chi_n^r$ (not shown). Their existence is likely associated with the previously described formation of bands of nearly degenerate eigenstates, and therefore they are highly susceptible to perturbations.

However, the main observation in Fig.~\ref{fig:fig8}(a) is that the behavior of both quantities differs significantly above the Anderson localization transition. The typical fidelity susceptibility, $\tilde{\chi}_\text{typ}^{r}$, exhibits a peak at a disorder strength close to, but lower than $W_2^* \approx 16.5$. This peak is narrow, causing $\tilde{\chi}_\text{typ}^{r}$ to quickly decay on the localized side. When $W \gtrsim 25$, $\tilde{\chi}_\text{typ}^{r}$ reverses its trend and starts to decrease with decreasing $\mu$. In contrast, the average fidelity susceptibility, $\tilde{\chi}_\text{av}^{r}$, develops a peak at a disorder strength that is slightly larger than $W_2^* \approx 16.5$. Moreover, the peak is broad, and the trend reversal is observed only for $W \gtrsim 40$.

In Fig.~\ref{fig:fig8}(b), we plot the ratio $\tilde{\chi}_\text{av}^{r} / \tilde{\chi}_\text{typ}^{r}$ versus the disorder strength $W$ for different frequency cutoffs $\mu$. The discrepancy between the average and typical fidelity susceptibilities decreases as $\mu$ increases. As we can observe, significant deviations from $\tilde{\chi}_\text{av}^{r} / \tilde{\chi}_\text{typ}^{r} = 1$ appear near the Anderson localization transition (indicated by a red vertical line) and in the localized regime for all considered $\mu$, at least for the fixed system size $V = 38^3$. In the inset of Fig.~\ref{fig:fig8}(b), we show how the ratio $\tilde{\chi}_\text{av}^{r} / \tilde{\chi}_\text{typ}^{r}$ varies with the frequency cutoff $\mu$ for three disorder strengths: $W=12.5$ (circles), $16.5$ (squares) and $25$ (triangles), which are located below, near and above the Anderson localization transition, respectively. We consider two system sizes: $V=28^3$ (blue symbols) and $38^3$ (black symbols). The numerical results are consistent with the ratio approaching unity (indicated by the red horizontal line) for all frequency cutoffs $\mu > \omega_\text{typ}$ in the large system size limit and disorder strengths $W \le W_2^*$. Conversely, in the localized regime, the results indicate that the ratio remains different from unity even for large frequency cutoffs, i.e., $\mu\gg \omega_\text{typ}$. This observation suggests that the lack of self averaging of the spectral functions (and so the fidelity susceptibilities) is not only related to the lowest frequency scales associated with $\omega_\text{typ}$.

%%%FIGURE
\begin{figure*}[!t]
\includegraphics[width=\textwidth]{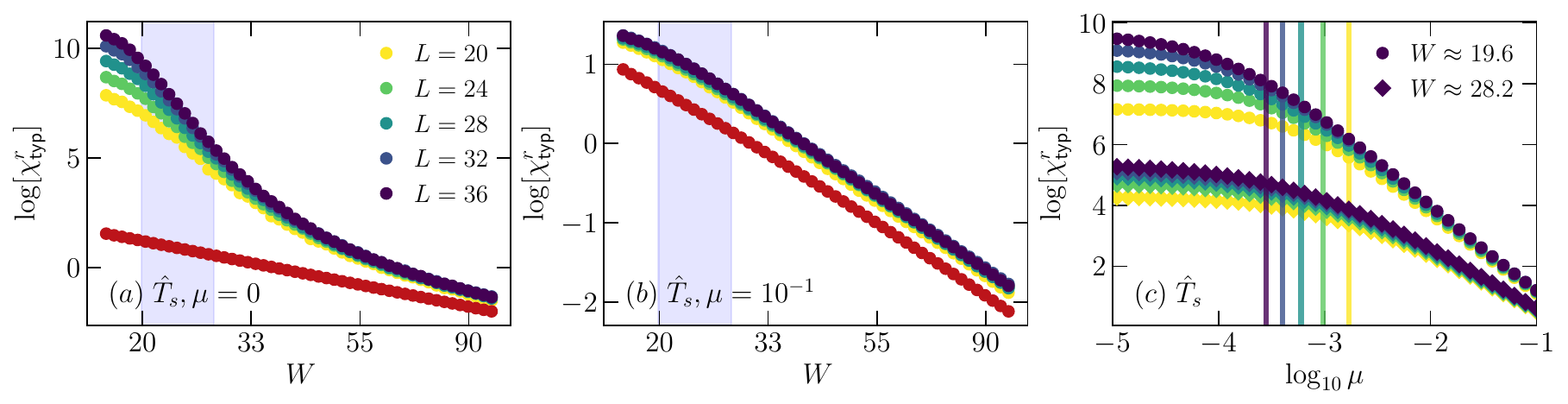}
\vspace{-0.2cm}
\caption{$\ln[\chi_\text{typ}^r]$ plotted vs [(a) and (b)] $W$ and (c) $\log_{10}\mu$.
The perturbation is $\hat{T}_{s}$. In (a) and (b), the frequency
cutoffs are $\mu = 0$ and $\mu =10^{-1}$, respectively. The upper curves show numerical results for the Anderson model, while the lower curve shows
the perturbation-theory prediction. The shaded region indicates the interval $W \in [20, 28]$, where $\tilde{\chi}_{\mathrm{typ}}^{\,r}$ exhibits a trend reversal. The horizontal axis is in logarithmic scale with tick marks at $\exp(3 + 0.5\, i)$,
where $i$ is an integer. Additionally, two disorder strengths are shown in (c): $W \approx 19.6$ and $W \approx 28.2$. The vertical line marks $\omega_{\mathrm{typ}}$.
}
\label{fig:fig10}
\end{figure*}
%%%FIGURE

The differences between the behavior of the typical and average fidelity susceptibilities in the localized regime can be understood intuitively.
For concreteness, we focus on a single observable, which is a normalized site occupation,  $\hat{n}_{i} = \sqrt{V}\,\hat{c}^\dagger_i \hat{c}_i - 1/\sqrt{V}$. A similar reasoning applies to other local observables. In the single-particle sector, we can write $\hat{c}^\dagger_{i} \hat{c}_{i} = \ket{i}\bra{i}$. Recall that in the localized regime, the single-particle energy eigenstates, $\ket{n}$ and $\ket{m}$, are exponentially localized, meaning that the probability amplitude of finding a particle drops off rapidly with distance from its localization center. Consequently, $\langle n | \hat{n}_{i} | m \rangle = \sqrt{V} \langle n | i \rangle \langle i | m \rangle$ is essentially zero for most $\ket{n}$ and $\ket{m}$, except for those with nearby localization centers. Such pairs, however, are rarely close in energy and, therefore, their contribution to $\tilde{\chi}^r_\text{typ}$ or $\tilde{\chi}^r_\text{av}$ is negligible. Nevertheless, the occasional resonances between sites lead to the hybridization of single-particle energy eigenstates and the formation of Mott pairs~\cite{Mott01061968,Ivanov_2012,Lecture_notes,Skvortsov_2022}, see Appendix~\ref{app:mott} for details.
When computing the fidelity susceptibility at small $\mu$, the resonances at low energy differences $|E_n-E_m|\lesssim \mu$ are strongly enhanced by small denominators. If these resonances are rare, they have little effect on $\chi_\text{typ}^r$, but can still lead to large values of $\chi_\text{av}^r$. 
This intuitive picture is further supported in Fig.~\ref{fig:fig9} of Appendix~\ref{sec:pbc}, which shows the distributions of fidelity susceptibilities.

\subsection{Different nonergodic regimes}

The above findings indicate the existence of three localized regimes, at least in finite systems. In the following (except for Sec.~\ref{subsub}), we focus on the unrescaled fidelity susceptibilities, i.e., $\chi^r_{\text{typ/av}}$ rather than $\tilde{\chi}^r_{\text{typ/av}}=\chi^r_{\text{typ/av}}\mu$. We note that the behavior of fidelity susceptibilities, similar to that described below, has been observed in interacting disordered spin models~\cite{PhysRevE.104.054105}.

When $W_2^* < W < W_3^*$, both fidelity susceptibilities diverge with decreasing $\mu$ or increasing $V$ at small enough $\mu$. Therefore, resonances occur at all energy scales and are not rare. 
Below, we argue that $W_3^* \approx 25$ for the system sizes under investigation.
Next, for $W_3^* < W < W_4^*$, the average fidelity susceptibility diverges under the same conditions, while the typical fidelity susceptibility grows more slowly than $\mu^{-1}$ with decreasing $\mu$ and becomes independent of $V$. This again implies that resonances are present at all energy scales, but they seem to be rare, since they affect $\chi^r_\text{av}$ more strongly than $\chi^r_\text{typ}$. 
Numerically, we estimate $W_4^* \approx 40$ for the system sizes under investigation (not shown). We cannot, however, make any claims about its thermodynamic limit value.
In general, the intermediate disorder regime with $W_2^* <W< W_4^*$ is analogous to the mixed phase space regime in classical systems, where regular trajectories/eigenstates with finite $\chi^r_n$ coexist with chaotic trajectories/eigenstates with divergent $\chi^r_n$~\cite{Lim_2024}.

Finally, when $W > W_4^*$, both fidelity susceptibilities grow more slowly than $\mu^{-1}$ with decreasing $\mu$, and become independent of $V$. This behavior is consistent with the adiabatic gauge potential (AGP), $\hat{\mathcal{A}}\ket{n(\lambda)} = i\hbar\,\partial_\lambda \ket{n(\lambda)}$, becoming a local operator with a finite norm. This AGP norm corresponds directly to the average fidelity susceptibility, $||\hat{\mathcal{A}}||^2 = \chi^r_{\text{av}}(\mu=0)$. Moreover, this locality can be interpreted as a signature of emergent integrability and well-defined local integrals of motion (LIOMs), as it implies local adiabatic connectivity between LIOMs and trivial integrals of motion at infinite disorder~\cite{PhysRevX.10.041017,Lim_2024}. We refer to this regime as the trivial insulator regime.

We show in Sec.~\ref{subsub} that $W_3^*$ appears stable in the large system-size limit, as it does not seem to flow toward $W_2^* \approx 16.5$ or toward infinite disorder strength.
We note that a similar transition was reported for the Gaussian Rosenzweig-Porter random matrix model in Ref.~\cite{Skvortsov_2022}. In contrast, the fate of $W_4^*$ is difficult to determine and hence we do not study it in detail here. It is possible that it drifts toward $W_3^* \approx 25$ in the large system-size limit, such that the system becomes a trivial insulator for all $W > W_3^*$.

\subsubsection{Perturbation theory}

To quantify the effects of resonances on the typical fidelity susceptibility, we compare numerical results with perturbation theory, which is expected to be valid at strong disorder, i.e., $W>W_3^*$. In this case, $\chi_{\rm typ}^r$ is expected to become $\mu$-independent at small $\mu$, which in turn implies that $\tilde{\chi}^r_{\rm typ} = \mu \chi^r_{\rm typ}$ vanishes at $\mu\to 0$. This occurs because the single-particle orbitals are strongly localized around individual sites, and resonances at $E_n \approx E_m$ are very rare, so they do not contribute significantly to the typical fidelity susceptibility. 

One can support this expectation by a simple perturbative argument performed around the limit of infinite disorder strength, where the zeroth-order eigenvalues, $\epsilon^{(0)}_n$, are given by the on-site potentials, and the zeroth-order eigenstates, $\ket{n^{(0)}}$, are perfectly localized on a single site. For the sublattice kinetic energy $\hat{T}_{s}$, the lowest-order approximation of the fidelity susceptibility of a single eigenstate $\ket{n}$ is given by
\begin{equation}
\label{eq:pert}
    \chi_n^r \approx 
    \sum_{m\neq n}
    \left(
        \frac{\omega_{mn}^{(0)}}{\left[\omega_{mn}^{(0)}\right]^2 + \mu^2}
    \right)^2
    \left|\langle n^{(0)} | \hat{T}_{s} | m^{(0)} \rangle\right|^2.
\end{equation}
Since the off-diagonal matrix elements of $\hat{T}_{s}$ simplify to $\langle n^{(0)} | \hat{T}_{s} | m^{(0)}\rangle = \sum_{n''}\delta_{m n''}$, with $n''$ denoting the next-nearest neighbors of $n$, the expression from Eq.~\eqref{eq:pert} can be rewritten as
\begin{equation}
    \chi_n^r \approx 
    \sum_{n''}
    \left(
        \frac{\omega_{n n''}^{(0)}}{
        \left[\omega_{n n''}^{(0)}\right]^2 + \mu^2}
    \right)^2 .
\end{equation}
The typical value of $|\omega_{nn''}|$ is $W$, so that the typical value of $\chi_n^r$ is independent of $\mu$ for small $\mu$ and proportional to $1/W^2$.

%%%FIGURE
\begin{figure}[!t]
\includegraphics[width=\columnwidth]{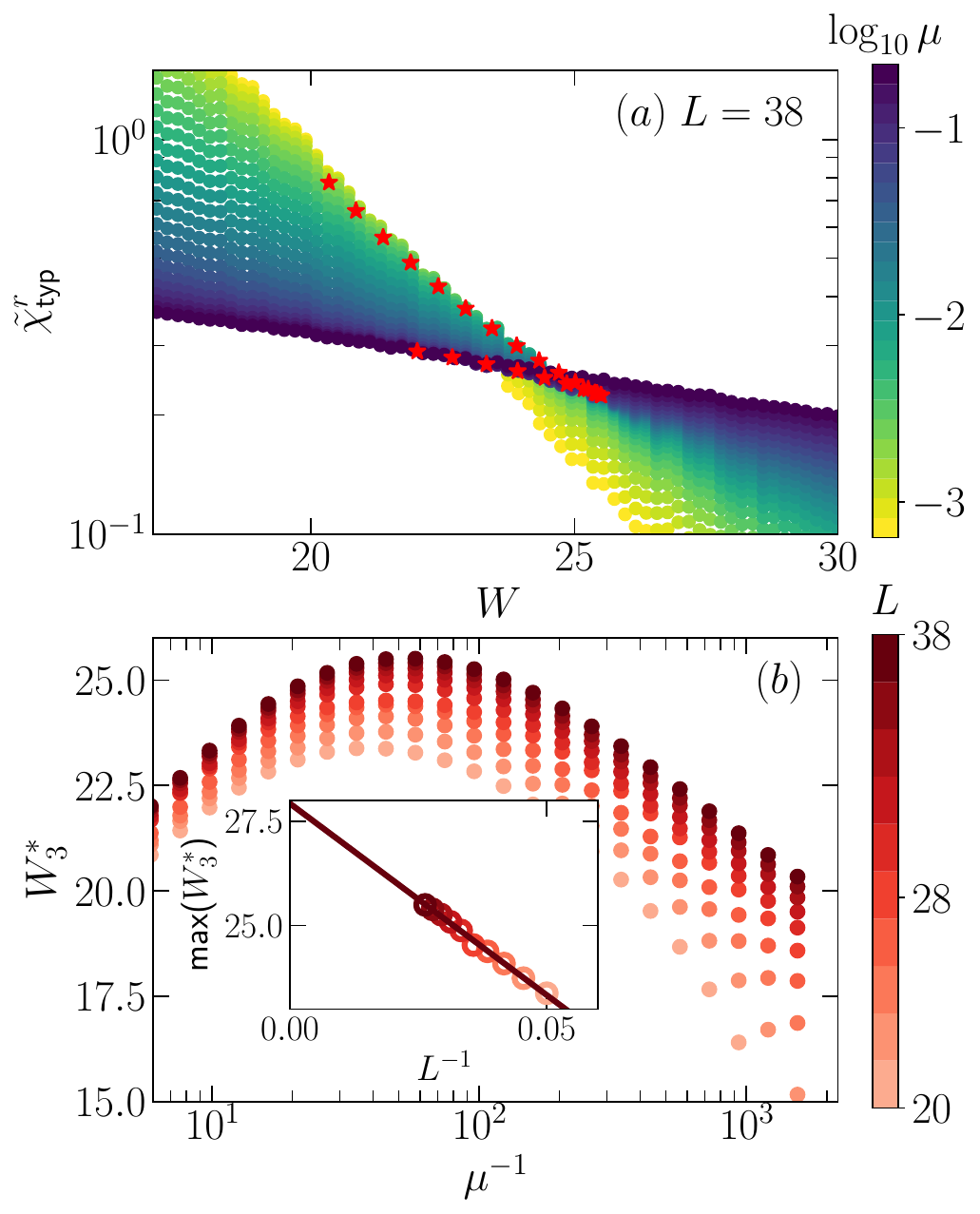}
\vspace{-0.2cm}
\caption{(a)~$\tilde{\chi}_\text{typ}^{r}$ of observable $\hat T_s$ vs $W$ in the regime where it changes its trend with $\mu$. Red stars indicate the crossovers, $W_3^*$, between curves for neighboring frequency cutoffs, $\mu$, which are equally spaced on a logarithmic scale. Results are shown for $L=38$. (b)~$W_3^*$ plotted versus $\mu^{-1}$ for $20 \leq L \leq 38$.  Each curve exhibits a maximum, and we determine its position from a second-order polynomial fit. This maximum appears to scale as $a_1+b_1 L^{-1}$ with $a_1 \approx 27.92$, as demonstrated in the inset.
}
\label{fig:fig11}
\end{figure}
%%%FIGURE

In Figs.~\ref{fig:fig10}(a) and \ref{fig:fig10}(b), we plot $\ln[\chi_\text{typ}^r]$ as a function of $W$ for $\mu = 0$ and $\mu = 10^{-1}$, respectively.  We consider $\hat{T}_{s}$ as the perturbation, and show results for different system sizes, as indicated in the legend of Fig.~\ref{fig:fig10}(a). The lower red curve is the perturbation theory prediction. 
The shaded region indicates the vicinity of $W_3^*$, see the discussion in the last paragraph of this section.
The horizontal axis is logarithmic. The perturbation-theory prediction appears to agree with the numerical results (up to a small vertical shift) for $W \gtrsim 25$ when the frequency cutoff is large, as in Fig.~\ref{fig:fig10}(b). Simultaneously, the agreement is reached only for $W\approx 90 \gg 25$ when the frequency cutoff is close to $\omega_{\text{typ}}$, as in Fig.~\ref{fig:fig10}(a). This indicates that even for $W \gtrsim 25$, which is already significantly larger than $W_2^\ast$, the system develops non-perturbative low-energy resonances. These resonances are not rare, as they influence the typical fidelity susceptibility.

We confirm this expectation in Fig.~\ref{fig:fig10}(c), where $\ln[\chi_\text{typ}^r]$ is plotted as a function of $\log_{10}\mu$ for two disorder strengths, $W \approx 19.6$ and $W \approx 28.2$. The vertical lines mark $\omega_{\text{typ}}$ for different $V$. When $W \lesssim 25$, although $\ln[\chi_\text{typ}^r]$ becomes independent of $\mu$ at small $\mu$, it acquires an additional system-size dependence when $\mu \lesssim \omega_{\rm typ}$. 
This indicates that $\omega_{\rm typ}$ is relevant for the behavior of the typical fidelity susceptibility, and that low-energy resonances can span the entire volume, making the insulator highly nontrivial. Nevertheless, when $W \gtrsim 25$, this system-size dependence becomes much weaker, suggesting a true insulating behavior, where low-energy resonances still occur but remain localized and insensitive to $V$.

\subsubsection{Finite-size scaling of $W_3^*$}
\label{subsub}

Finally, we investigate whether the described crossover at $W_3^\ast$ shifts toward $W_2^\ast$, saturates at a value greater than $W_2^\ast$, or diverges toward infinite disorder strength in the thermodynamic limit. For this reason, we determine $W_3^*$ from the crossing points of $\tilde{\chi}^r_\text{typ}$ (again rescaled) for neighboring $\mu$, which are equally spaced on the logarithmic scale. These crossing points are marked with red stars for $L=38$ in Fig.~\ref{fig:fig11}(a) and plotted versus $\mu^{-1}$ for different $L$ in Fig.~\ref{fig:fig11}(b). Initially, $W_3^*$ increases with $\mu^{-1}$, but then it exhibits a maximum and starts to decay as $\mu \rightarrow \omega_\text{typ}$. Therefore, it clearly depends not only on the system size $L$ but also on the frequency cutoff $\mu$. Nevertheless, for all justifiable choices of $\mu$, $W_3^*$ does not converge to $W_2^*$ in the large system-size limit. It seems reasonable to define the edge of the intermediate regime, in which the behavior of $\tilde{\chi}^r_\text{typ}$ is dominated by the Mott pairs, by the disorder strength corresponding to the maximal $W_3^*$. The latter is plotted against $L^{-1}$ in the inset of Fig.~\ref{fig:fig11}(b), and it appears to closely follow $\text{max}(W_3^*) = a_1 + b_1 L^{-1}$ with $a_1 \approx 27.9$.

%%%%%%%%%%%%%%%%%%%%%%%%%%%%%%%%%%%%%%%%%

%%%%%%%%%%%%%%%%%%%%%%%%%%%%%%%%%%%%%%%%%
\section{Conclusions}
\label{sec_con}

In this work, we studied the behavior of fidelity susceptibilities as the system evolves through the chaotic and non-chaotic regimes of the 3D Anderson model.
This model hosts the crossover at weak disorder and the localization transition at moderate disorder.
The weak-disorder crossover occurs between localization in quasimomentum space and single-particle quantum chaos, for which the critical disorder strength $W_1^*$ vanishes in the infinite system-size limit $V\rightarrow\infty$.
The Anderson localization transition occurs between single-particle quantum chaos and localization in position space at the critical disorder strength $W_2^*\approx 16.5$. 

The main result of our work is that, as a function of disorder, the fidelity susceptibility indeed exhibits two peaks at disorder strength $W_1^*$ and $W_2^*$, which correspond to the weak-disorder crossover and the localization transition, respectively.
From the position of the first peak, we found that $W_1^* \sim 1/\sqrt{V}$, in agreement with the Fermi golden rule based criterion for the crossover.
From the position of the second peak, we recovered the well-established prediction for the localization transition, $W_2^* \approx 16.5$~\cite{Slevin_2018}, with a relative error of only a few percent. 

We also considered the scaling of fidelity susceptibilities with system size, both in the regularized form, $\tilde{\chi}_\text{typ}^r$, that contains the frequency cutoff $\mu$, and in the unregularized form, $\tilde{\chi}_\text{typ}$. 
We observed that the fidelity susceptibility is maximally divergent at the weak-disorder crossover. We argued that the later can be described in terms of fading ergodicity~\cite{Kliczkowski_2024, PhysRevB.111.184203}, which is confirmed by the behavior of the spectral function $|f(\omega)|^2$. 

Interestingly, while the peak of the fidelity susceptibilities scales polynomially with the system size at the localization transition, the corresponding scaling exponent is sub-maximal. 
We explained this behavior by relating this exponent to the fractal dimension $d_2$ of the critical wavefunctions at the Anderson localization transition.
This analysis was complemented by exploring the behavior of the spectral function $|f(\omega)|^2$ across the localization transition.

Finally, we addressed the scaling of fidelity susceptibility $\tilde{\chi}_\text{typ}^r$ with the frequency cutoff $\mu$. Specifically, we discussed under what conditions a generally two-parameter scaling, governed by the frequency cutoff $\mu$ and the system size $V$, simplifies to a one-parameter scaling.
We also explored the differences between the typical and average values of fidelity susceptibility. We find that these differences are negligible over a broad range of disorder strengths, but they become significant near the critical disorder strength $W_2^*$, and in the localized regime at $W > W_2^*$. 
In particular, we find evidence of two distinct regimes of nonergodic behavior within the localized phase, and the crossover between them, emerging at the disorder strength $W_3^* > W_2^*$, is defined by the scale invariant point of $\tilde{\chi}_\text{typ}^r$.
These two regimes may be referred to as a trivial insulator [at $W>W_3^*$] and a nontrivial Anderson insulator [at $W_2^* < W < W_3^*$], and a more thorough characterization of their properties is left for future work.

%%%%%%%%%%%%%%%%%%%%%%%%%%%%%%%%%%%%%%%%%

\acknowledgements 
We acknowledge discussions with D.~Sels, M.~Mierzejewski, R.~Świętek, M.~Kliczkowski, M.~Hopjan and R.~Sharipov. Numerical studies in this work have been carried out using resources provided by the~Wroclaw Centre for Networking and Supercomputing, Grant No. 579 (P.~T., P.~{\L}.). 
L.V. acknowledges support from the Slovenian Research and Innovation Agency (ARIS), Research core funding Grants No.~P1-0044, N1-0273 and J1-50005.
A.P. was supported by NSF grant DMR-2412542 and AFOSR grant FA9550-21-1-0342.

\appendix

\section*{DATA AVAILABILITY}
The data that support the findings of this article are openly available~\cite{data_repository}.

%%%%%%%%%%%%%%%%%%%%%%%%%%%%%%%%%%%%%%%%%
\section{Phenomenological model} \label{sec:app_phenomenological}

In the main text, we discussed two different scenarios for the emergence of the peak of fidelity susceptibility.
It turns out that it is possible to merge these two scenarios into a single framework with the help of the phenomenological model proposed in Ref.~\cite{PhysRevLett.127.230603}, at least for the standard integrability-breaking transitions. In that work, the authors assumed that at the integrable point, the low-frequency part of the spectral function of observable $\hat O$ can be represented as a sum of Drude-like peaks:
\begin{equation}
\label{eq:Drude}
|f(\omega)|^2 \sim \sum_{j=1}^N D_j \delta(\omega),
\end{equation}
where $\delta(\omega)$ is the Dirac delta, $N$ denotes the total number of LIOMs, while the coefficients can be calculated as $D_j=\langle\hat{O}\hat{I}_{j}\rangle^2/\langle\hat{I}_j\hat{I}_j\rangle$ with $\langle\hat{O}\hat{I}_{j}\rangle=\frac{1}{Z}\sum_{n} \langle n|\hat{O} \hat{I}_{j} | n\rangle$ standing for the Hilbert-Schmidt product of observable $\hat{O}$ and LIOM $\hat{I}_{j}$. The key assumption of Ref.~\cite{PhysRevLett.127.230603} is that LIOMs $\hat{I}_{j}$ acquire finite relaxation rates $\Gamma_j$ upon the introduction of weak integrability breaking, leading to the broadening of Drude-like peaks:
\begin{equation}
\label{eq:eq_envelope}
    |f(\omega)|^2 \sim \sum_{j=1}^{N} D_j \frac{1}{\pi}\frac{\Gamma_j}{\omega^2+\Gamma_j^2}.
\end{equation}
Note that the approximate LIOMs may not correspond to those of the parent integrable model~\cite{Pawlowski_2025}. The only consequence for the argument presented below is that the set of $\hat{I}_j$ may depend on the integrability-breaking parameter $\Delta$. A similar framework may apply to other transitions, in which LIOMs are not clearly defined. In such cases, the sum may run over eigenstates of the unperturbed model that acquire finite relaxation rates when the perturbation is introduced. Therefore, for simplicity, we refer to each contribution to the sum as a slow mode.

%%%FIGURE
\begin{figure}[!t]
\includegraphics[width=\columnwidth]{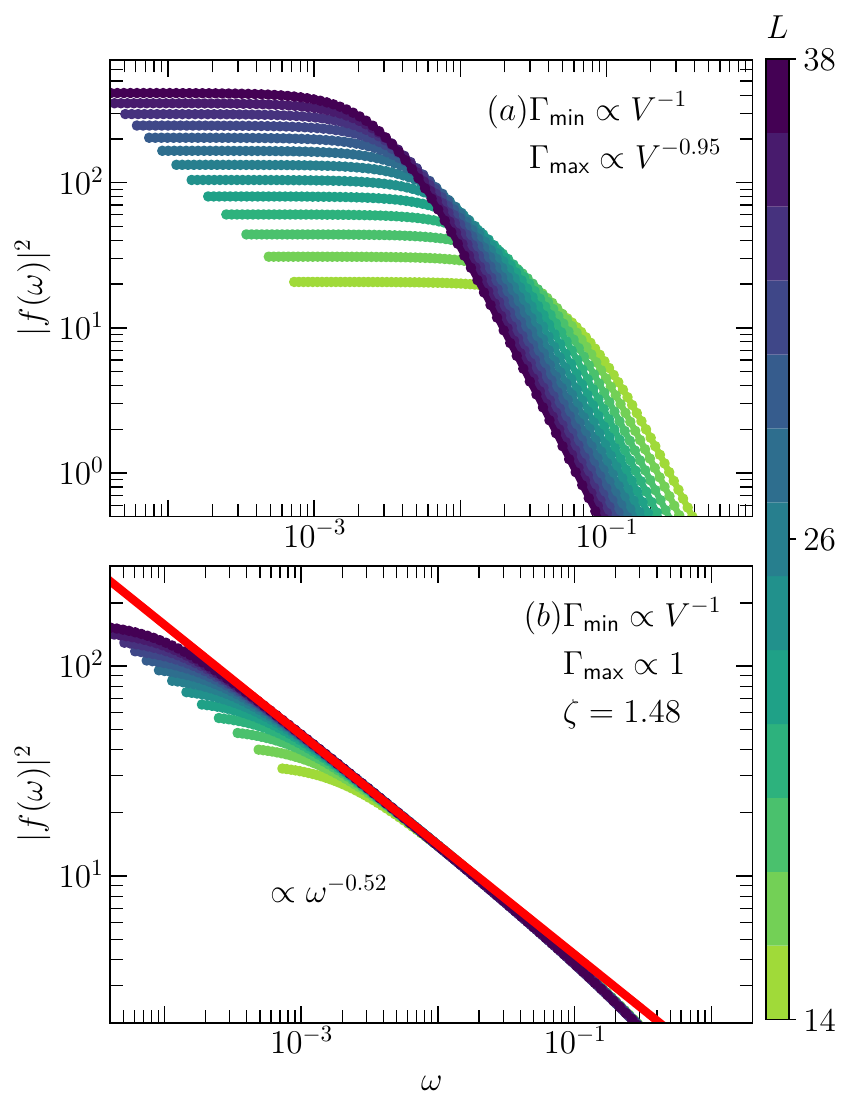}
\vspace{-0.2cm}
\caption{The spectral functions from the phenomenological model, which correspond to the two limiting scenarios:
(a) fading ergodicity and
(b) slowing down of polynomial relaxation of observables.
The relevant scaling parameters are indicated in each panel and discussed in the text.
The red curve in panel (b) shows the least-squares fit to the data with $V = 38^3$, using $b\,\omega^{-a}$.
We obtain $a \approx 0.52$.
}
\label{fig:figa00}
\end{figure}
%%%FIGURE

Let $N \gg 1$, i.e., we assume that the observable $\hat{O}$ is coupled to many slow modes. We also assume that there are no correlations between $D_j$ and $\Gamma_j$. Then, two simplifications follow. First, the coefficients $D_j$ can be replaced by their mean: $D_j\rightarrow \frac{1}{N}\sum_{j}D_{j}=\frac{D_0}{N}$, with $D_0$ bounded from above by the Hilbert-Schmidt norm of $\hat{O}$. Second, the sum over slow modes can be replaced by the integral over relaxation rates: $\frac{1}{N}\sum_{j}\rightarrow A\int_{\Gamma_\text{min}}^{\Gamma_\text{max}} d\Gamma\; \Gamma^{\zeta-2}$, where $A$ is the normalization factor and $\zeta$ controls the distribution width. Finally,
\begin{equation}
\label{eq:explanation}
    |f(\omega)|^2 \sim \frac{AD_0}{\pi} \int_{\Gamma_\text{min}}^{\Gamma_\text{max}} d\Gamma\; \frac{\Gamma^{\zeta+1}}{\Gamma^2+\omega^2}.
\end{equation}
An important outcome of Ref.~\cite{PhysRevLett.127.230603} is that the envelope function behaves as a polynomial, $\propto\omega^{-(2-\zeta)}$, in the range $\Gamma_{\rm min}\lesssim \omega \lesssim \Gamma_\text{max}$. Moreover, this behavior is independent of the particular choice of broadening [which is selected to be Lorentzian in Eqs.~\eqref{eq:eq_envelope} and~\eqref{eq:explanation}].

To proceed, we note that the relaxation rates of slow modes, $\Gamma_j$, generally depend on both the system volume, $V$, and the strength of perturbation, $\Delta$. The transition point can be defined as the condensation point, where the slowest relaxation rates approach the Heisenberg scale, i.e., $\Gamma_\text{min}\sim \omega_H\propto Z^{-1}$ ~\cite{PhysRevE.102.062144}.
Note that the inverse of $\Gamma_\text{min}$ defines the longest relaxation time and is often identified with the Thouless time~\cite{Edwards_1972, Gharibyan_2018, Schiulaz_2019, PhysRevE.102.062144, Suntajs_2021}. 

If $\Gamma_\text{max} \gg \Gamma_\text{min}$, a transient frequency regime $\Gamma_\text{min} \lesssim \omega \lesssim \Gamma_{\rm max}$ inevitably emerges, in which, under very general assumptions and for $N \gg 1$, the spectral function $|f(\omega)|^2$ exhibits a power-law scaling $\propto\omega^{-(2-\zeta)}$, with the phenomenological parameter $\zeta$ determined by the details of the distribution of $\Gamma$. Therefore, if $\Gamma_\text{max} \gg \Gamma_\text{min}$, the phenomenological model is consistent with the second limiting scenario, namely, the slowing down of polynomial relaxations of observables. This is confirmed in Fig.~\ref{fig:figa00}(b). The parameters of the model have been selected so that the results for the 3D Anderson model near the localization transition from Fig.~\ref{fig:fig3}(b) are best reproduced.

If $\Gamma_\text{min} \approx \Gamma_\text{max}$ (or if $\hat{O}$ couples to just a few slow modes), slow modes freeze at similar perturbation strengths. As a result, the gradual softening of fluctuations occurs within a similar range of $\Delta$ for all of them, leading to the formation of a system-size-dependent plateau in $|f(\omega)|^2$. This behavior is confirmed in Fig.~\ref{fig:figa00}(a). Again, the parameters of the model have been selected so that the results for the 3D Anderson model near the weak-disorder crossover from Fig.~\ref{fig:fig3}(a) are best reproduced. It is important to emphasize that this limiting scenario depends on the choice of broadening. There is no guarantee that the contributions to $|f(\omega)|^2$ are functions that exhibit system-size-dependent plateaus near the transition point, as Lorentzians do. Consequently, fading ergodicity may or may not occur, and $|f(\omega)|^2$ may or may not exhibit system-size dependence. 
%%%%%%%%%%%%%%%%%%%%%%%%%%%%%%%%%%%%%%%%%

\section{Observables in the Anderson model}
\label{app:sublattice}

\begin{figure}[!t]
\includegraphics[width=0.85\columnwidth]{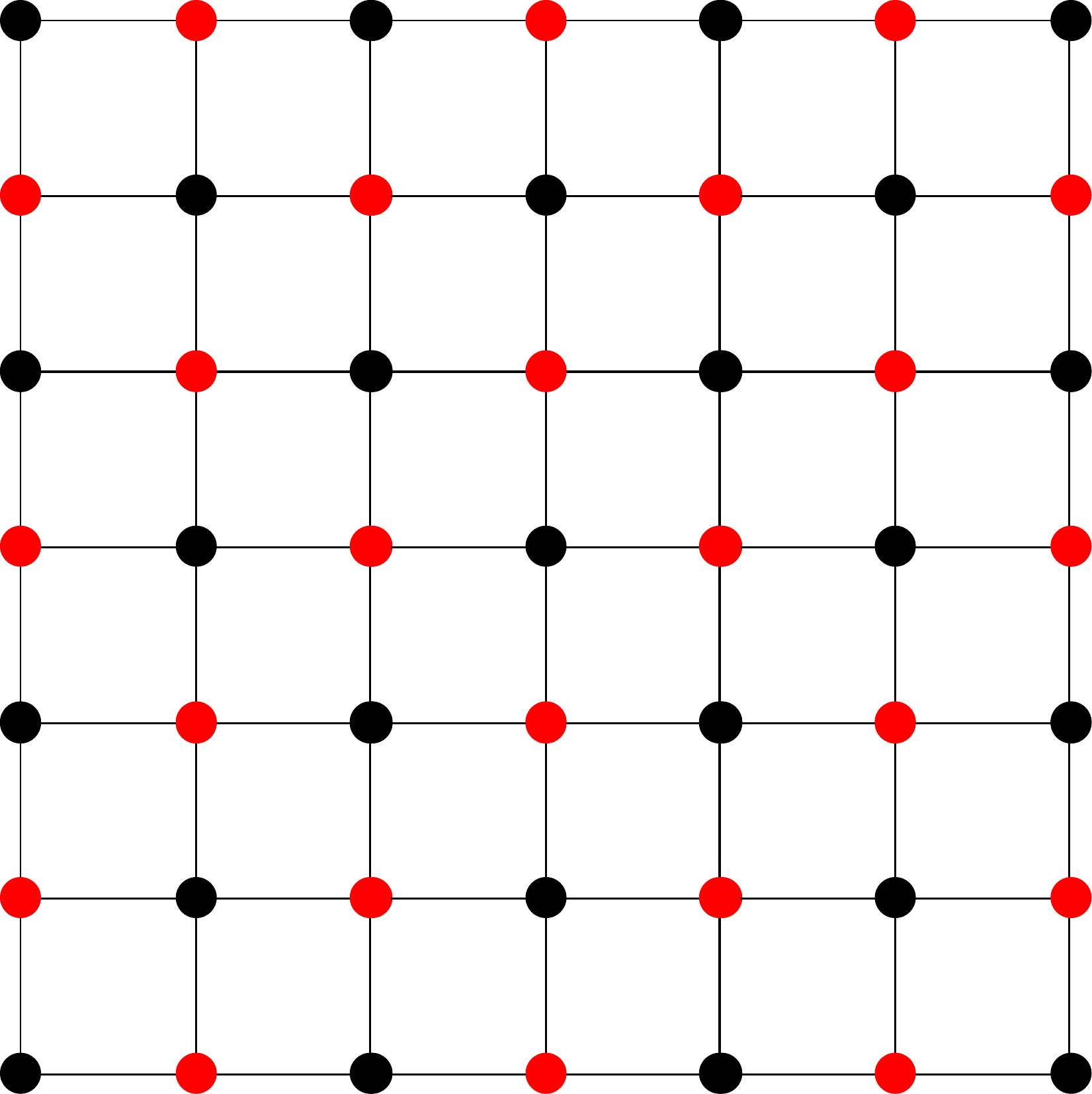}
\vspace{-0.2cm}
\caption{A cross-section of a cubic lattice with $V=7^3$. Sets of next-nearest neighbors $\langle\langle i,j \rangle\rangle_1$ and $\langle\langle i,j \rangle\rangle_2$ are marked with black and red, respectively.}
\label{fig:fig0a}
\end{figure}

In Eq.~\eqref{def_Ts} of the main text, we introduced the sublattice kinetic energy, 
$\hat{T}_s$, which divides the cubic lattice into two disjoint cubic sublattices indexed by $\alpha\in\{1,2\}$. Each set $\langle\langle i,j \rangle\rangle_\alpha$ represents a set of nearest neighbors within a sublattice (or next-nearest neighbors within the full lattice) with a hopping amplitude $t_\alpha = -1/\alpha$. These two sets are shown in different colors in a cross-section of a cubic lattice with $V = 7^3$ in Fig.~\ref{fig:fig0a}.

We also note that the behavior of the autocorrelation function $\langle \hat{n}(t)\hat{n}\rangle$ is expected to be consistent with that studied in the transport context of particle-number-conserving systems, i.e., $\langle \hat{n}_{i}(t)\hat{n}_{i}\rangle$, once averaged over disorder realizations. This follows from the relation
\begin{equation}
\begin{split}
    &\langle\!\langle \hat{n}(t)\hat{n}\rangle\!\rangle_\text{dis} \sim \iint d^3r\, d^3r^{\,\prime}\, g(\vec{r}) g(\vec{r}^{\,\prime}) \, \langle\!\langle \hat{n}(\vec{r},t)\hat{n}(\vec{r}^{\,\prime},0)\rangle\!\rangle_\text{dis} \\
    & \sim \iiint d^3r\, d^3r^{\,\prime}d^3k\, g(\vec{r}) g(\vec{r}^{\,\prime})e^{i\vec{k}(\vec{r}-\vec{r}^{\,\prime})} \, \langle \hat{n}(\vec{k},t)\hat{n}(\vec{k},0)\rangle,
\end{split}
\end{equation}
where we have adopted the continuum representation, replacing discrete site indices $i,j$ with spatial coordinates $\vec{r},\vec{r}^{\,\prime}$, lattice sums with spatial integrals, and site-dependent quantities $r_i$ with functions $g(\vec{r})$. Additionally, $\langle\dots\rangle_\text{dis}$ denotes averaging over Hamiltonian realizations. Using $\int d^3r\,g(\vec{r})e^{i\vec{k}\vec{r}}=g(\vec{k})$ and $g(\vec{k})^*=g(-\vec{k})$ for real $g(\vec{r})$, we obtain
\begin{equation}
    \langle\!\langle \hat{n}(t)\hat{n}\rangle\!\rangle_\text{dis}\sim\int d^3k\,|g(\vec{k})|^2 \langle \hat{n}(\vec{k},t)\hat{n}(\vec{k},0)\rangle,
\end{equation}
which has contribution from autocorrelation functions of many $\vec{k}$ modes. Note that for $g(\vec{r})=1$, only the autocorrelation function of the slowest mode with $\vec{k}=0$ is present. The latter corresponds to the total particle number, whose dynamics is frozen.

\section{Boundary disorder}
\label{app:boundary}

When considering the sublattice kinetic energy, $\hat{T}_{s}$, we add a small boundary term to the 3D Anderson model of the form
\begin{equation}
    \hat{H}_\text{edge}=\sum_{i\in\text{edge}} \tilde{\varepsilon}_i \hat{c}_{i}^\dagger \hat{c}_{i},
\end{equation}
where the sum runs over all sites located at the edge of the system, while $\tilde{\varepsilon}_i$ are i.i.d.~random variables uniformly sampled from $[-0.15,0.15]$. This boundary term does not affect the fidelity susceptibilities for $W>0.3$. At the same time, it leads to a more consistent system-size dependence of the peak marking the weak-disorder crossover at $W=W_1^*$. Without this term, the peak is still visible. However, for certain system sizes, the slope of its left shoulder changes abruptly. These are the same system sizes for which the average gap ratio saturates at a different value in the zero-disorder limit, see Fig.~\ref{fig:fig1a} in App.~\ref{app:spectral}.

\section{Spectral properties}
\label{app:spectral}

\begin{figure}[!t]
\includegraphics[width=\columnwidth]{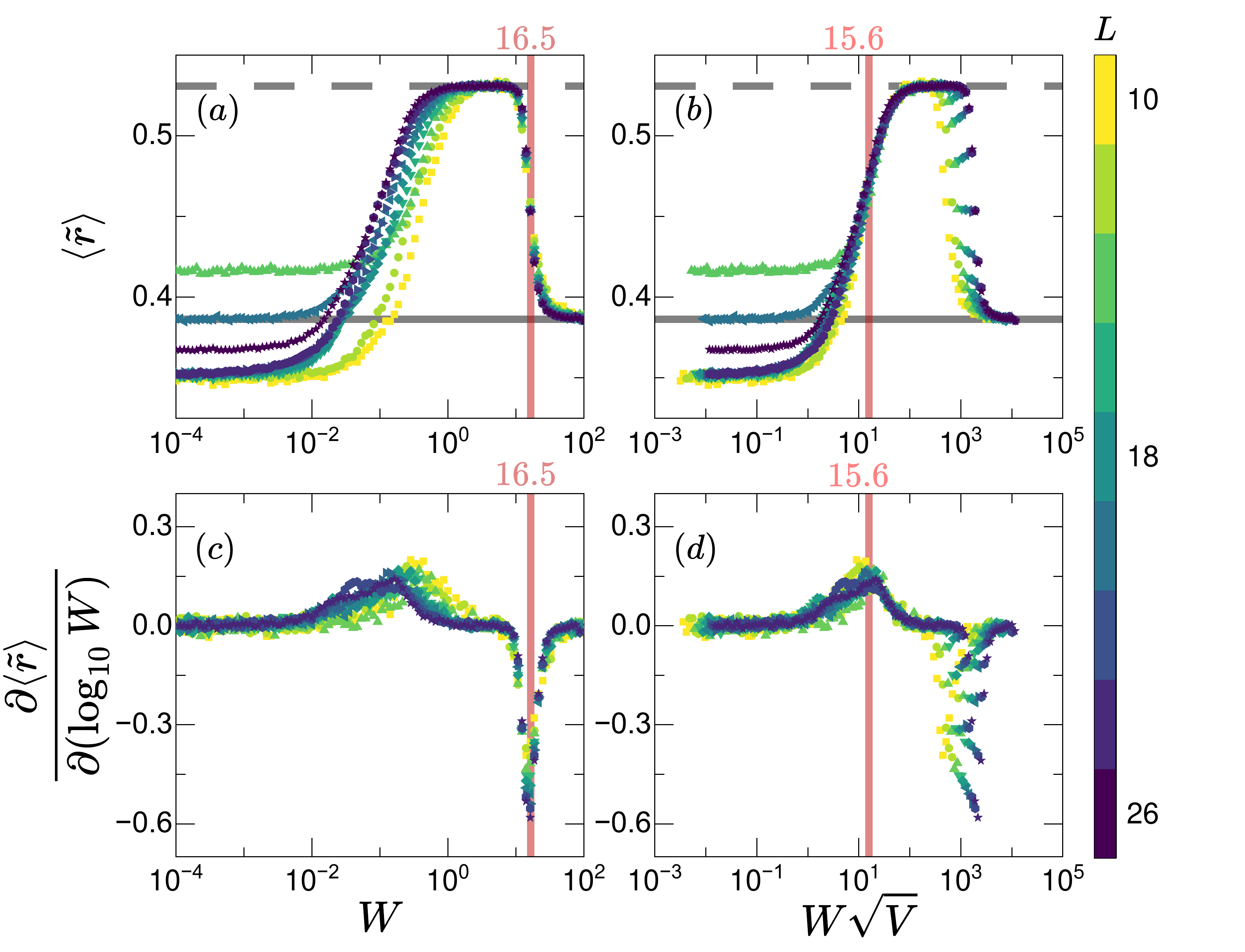}
\vspace{-0.2cm}
\caption{The average ratio $\langle{\tilde{r}}\rangle$ plotted versus (a)~$W$ and (b)~$W\sqrt{V}$. The derivative $\partial \langle \tilde{r} \rangle / \partial \log_{10} W$ plotted versus (c)~$W$ and (d)~$W\sqrt{V}$. Solid (dashed) horizontal lines in (a,b) mark $\tilde{r}_\text{P} \approx 0.386$ ($\tilde{r}_\text{GOE} \approx 0.5307$). Solid vertical lines in all panels indicate the points of the most rapid change of $\langle \tilde{r} \rangle$, where $\partial \langle \tilde{r} \rangle / \partial \log_{10} W$ is maximal. All numerical results were averaged over all single-particle energy eigenstates and subsequently over $40$ disorder realizations. }
\label{fig:fig1a}
\end{figure}

Spectral statistics are among the most widely used measures of quantum chaos and ergodicity.
Here, we focus on the gap ratio~\cite{PhysRevB.75.155111}: 
\begin{equation}
    \tilde{r}_{n} = \frac{\min[\delta_{n+1},\,\delta_{n}]}{\max[\delta_{n+1},\,\delta_{n}]},
\end{equation}
where $\delta_n = E_{n} - E_{n-1}$ denotes the energy spacing (gap) between neighboring levels $\ket{n}$ and $\ket{n-1}$. Note that $0 \le \tilde{r}_{n} \le 1$, provided that there are no multifold degeneracies in the system.
It has been demonstrated~\cite{Tarquini17, Suntajs_2021} that in the chaotic regimes, including the single-particle chaotic regime in the 3D Anderson model, when this ratio is averaged over many energy eigenstates, it agrees with the prediction of the Gaussian Orthogonal Ensemble, i.e., $\tilde{r}_\text{GOE} \approx 0.5307$~\cite{PhysRevLett.110.084101}. 
In the non-chaotic regimes, this average ratio, $\langle{\tilde{r}}\rangle$, usually agrees with the Poisson ensemble prediction, i.e., $\tilde{r}_\text{P} \approx 0.386$~\cite{PhysRevLett.110.084101}.

\begin{figure}[!t]
\includegraphics[width=\columnwidth]{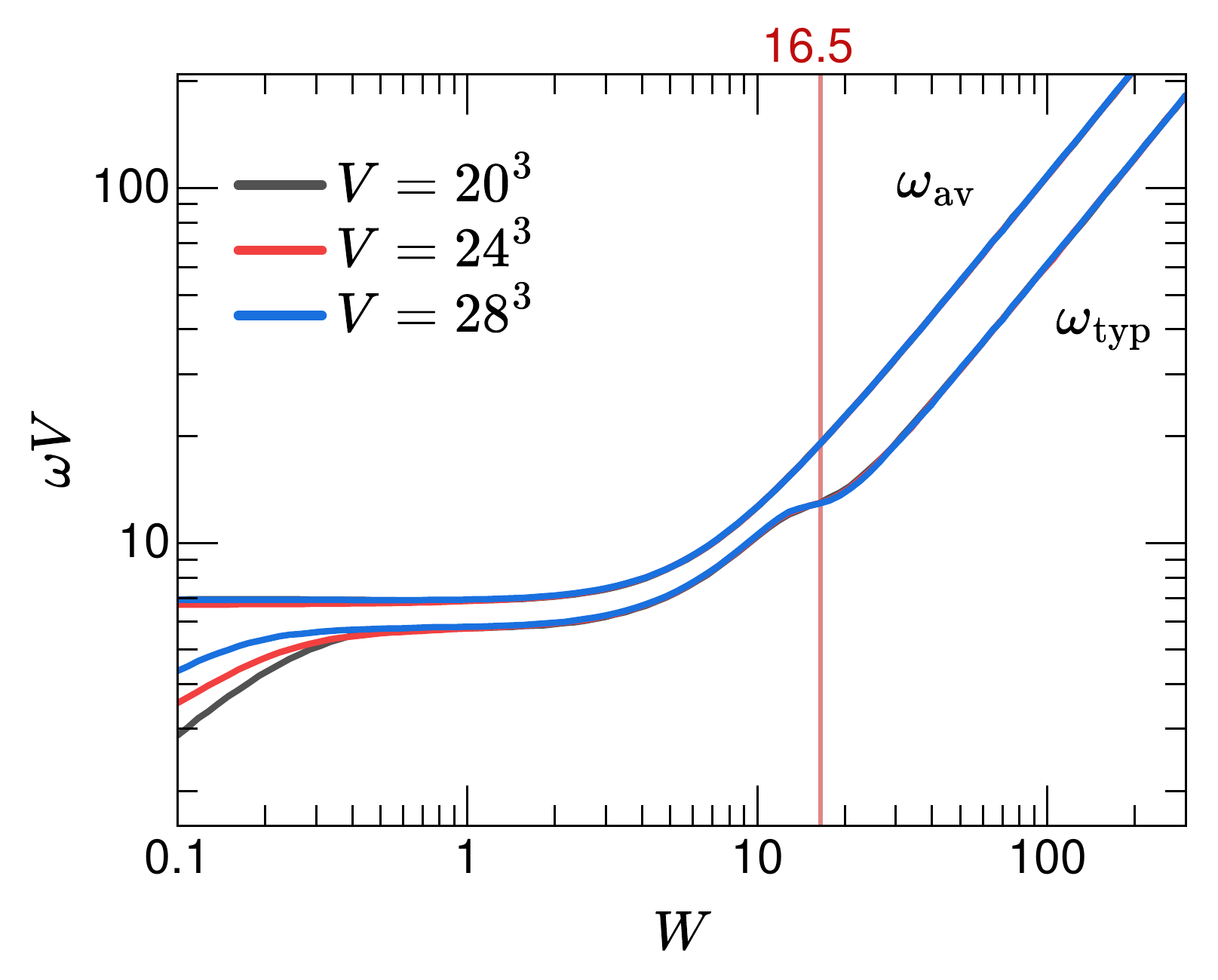}
\vspace{-0.2cm}
\caption{The typical and average level spacings, $\omega_\text{typ}$ and $\omega_\text{av}$, plotted versus the disorder strength, $W$. The vertical axis is rescaled with the system size $V$. Three system sizes are considered: $V=20^3, 24^3$ and $28^3$. The Anderson localization transition at $W\approx 16.5$ is marked with the red vertical line.}
\label{fig:fig2a}
\end{figure}

In Figs.~\ref{fig:fig1a}(a) and~\ref{fig:fig1a}(b), we plot $\langle \tilde{r} \rangle$ as a function of $W$ and $W\sqrt{V}$, respectively. 
The ratio is averaged over all single-particle energy eigenstates and subsequently over $40$ disorder realizations. 
In agreement with expectations, $\langle \tilde{r} \rangle$ remains close to $\tilde{r}_\text{GOE} \approx 0.5307$ (dashed horizontal line) for moderate disorders and drops towards $\tilde{r}_\text{P} \approx 0.386$ (solid horizontal line) for large disorders.
The change in behavior of $\langle \tilde{r} \rangle$ seems to occur at a fixed value of $W$.
Simultaneously, for low disorders and sufficiently large system sizes, $\langle \tilde{r} \rangle$ drops below the Poisson ensemble prediction.
This behavior seems to occur below a fixed value of $W\sqrt{V}$.
It indicates, together with more pronounced finite-size effects, the formation of subbands of nearly degenerate states. Such formation is not unexpected, as the system is close to translational invariance and quasimomentum is nearly conserved. We confirm this expectation in Fig.~\ref{fig:fig2a}, where we plot the typical level spacing, $\omega_\text{typ}$, and compare it to the average level spacing, $\omega_\text{av}$, for three different system sizes $V=20^3, 24^3$ and $28^3$. It is apparent that $\omega_\text{typ}\sim\omega_\text{av}$ for (almost) all disorder strengths $W$. Only when $W\ll 1$ does the typical level spacing $\omega_\text{typ}^{-1}$ begins to grow faster than linearly with the system size $V$. Moreover, $\omega_\text{typ}$ develops a shoulder or plateau near $W\approx 16.5$, which is not reflected in $\omega_\text{av}$, and it signals the Anderson localization transition.

Finally, we recognize the eigenstate transitions when the average ratio, $\langle \tilde{r} \rangle$, exhibits the most rapid change, i.e., its derivative, $\partial \langle \tilde{r} \rangle / \partial \log_{10} W$, is maximal. We plot this derivative as a function of $W$ and $W\sqrt{V}$ in Figs.~\ref{fig:fig1a}(c) and~\ref{fig:fig1a}(d), respectively. It was computed numerically using the two-point formula, and it exhibits two peaks. 
The first peak is broad and noisy, making its exact position difficult to determine, but it clearly scales as $1/\sqrt{V}$, consistent with the critical disorder strength of the weak-disorder crossover. 
Despite this, it emerges at lower disorder strengths than the peak of the fidelity susceptibility, for which $W_1^* \approx 41/\sqrt{V}$ (see the main text). 
The second peak is narrower and more sharply defined, with its position close to the critical disorder strength of the Anderson localization transition, $W_2^* \approx 16.5$.

\section{Spectral function under disorder variation}
\label{app:spectral}

\begin{figure}[!t]
\includegraphics[width=\columnwidth]{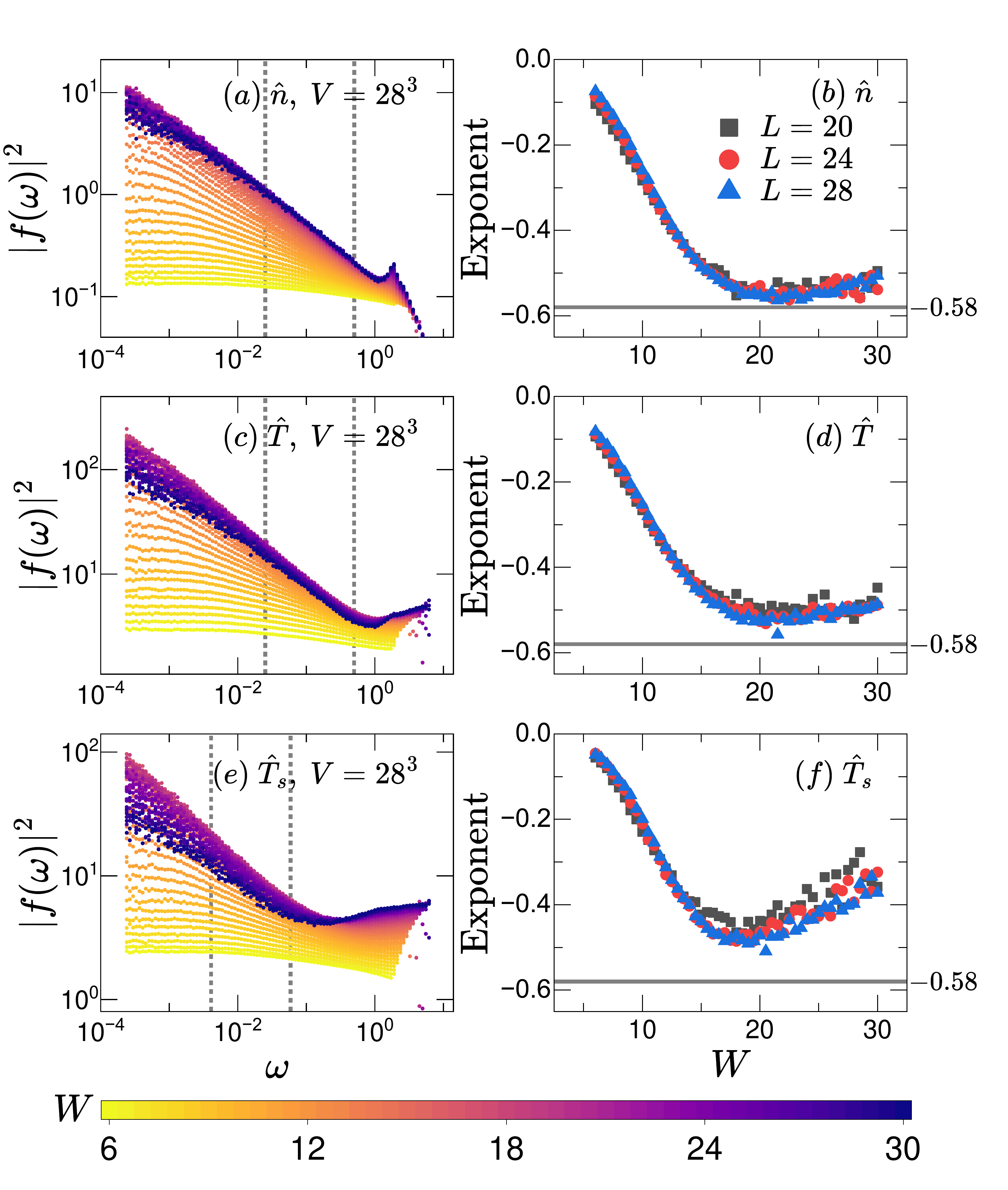}
\vspace{-0.2cm}
\caption{Spectral function $|f(\omega)|^2$ plotted versus $\omega$ for (a) $\hat{n}$, (b) $\hat{T}$ and (c) $\hat{T}_{s}$. We consider a single system size, $V = 28^3$, and several disorder strengths, $W \in [6,30]$. All numerical results were established from $20\%$ of single-particle energy eigenstates in the middle of the spectrum, and then averaged over $20$ realizations of disorder. The dashed lines in panels (a), (c), and (e) indicate the moderate-$\omega$ regime, where the least-squares fit of $b\omega^{-a}$ is performed. The resulting exponent $-a$ is plotted as a function of $W$ in panels (b), (d), and (f) for $\hat{n}$, $\hat{T}$, and $\hat{T}_s$, respectively. We focus on three system sizes: $V=20^3$, $24^3$, and $28^3$. The horizontal line marks the prediction $a=1-d_2\approx 0.58$ for $\hat{n}$.}
\label{fig:fig3a}
\end{figure}

It is interesting to examine the behavior of the spectral function $|f(\omega)|^2$, see Eq.~\eqref{eq:spectral} in the main text, as the Anderson localization transition is approached. In Figs.~\ref{fig:fig3a}(a), \ref{fig:fig3a}(c) and \ref{fig:fig3a}(e), we plot $|f(\omega)|^2$ for perturbations $\hat{n}$, $\hat{T}$ and $\hat{T}_{s}$, respectively. The minimal considered energy mismatch, $\omega$, corresponds to the typical level spacing, $\omega_\text{typ}$. We show $|f(\omega)|^2$ for a single system size, $V = 28^3$, and several disorder strengths, $W \in [6,30]$. They are established from $20\%$ of the single-particle energy eigenstates in the middle of the spectrum, and then averaged over $20$ realizations of disorder.

In agreement with expectations, the spectral function for all perturbations develops a plateau in the low-$\omega$ regime, which disappears around $W \approx 16.5$, i.e., close to the Anderson localization transition. This indicates that the longest relaxation time, $\tau_\text{max}$, becomes comparable to the Heisenberg time, $\tau_\text{H} = \omega_\text{typ}^{-1}$. Recall that local observables are expected to rapidly relax towards non-thermal values deep in the localized regime, so that the spectral function may either redevelop a plateau or open a gap (i.e., $|f(\omega)|^2 \rightarrow 0$ as $\omega \rightarrow 0$) when $W\gg 16.5$.

Simultaneously, the plateau evolves into a polynomial tail, in which $|f(\omega)|^2 \sim \omega^{a}$, at larger $\omega$.
The exponent $a$ decreases as the Anderson localization transition is approached and then becomes weakly dependent on $W$ in the localized regime (at least for not too large $W$). In our phenomenological model, this behavior is reproduced by the distribution of relaxation times that broadens (i.e., $\zeta$ decreases) as the disorder strength approaches the Anderson localization transition, and then remains nearly unchanged or slowly narrows (i.e., $\zeta$ increases) in the localized regime. 

Finally, we complement the study by performing the least-squares fit of $b\omega^{-a}$ within the moderate-$\omega$ regime, which is indicated by the dotted vertical lines in Figs.~\ref{fig:fig3a}(a), \ref{fig:fig3a}(c) and \ref{fig:fig3a}(e). The exponent $-a$ is then plotted versus $W$ for $\hat{n}$, $\hat{T}$ and $\hat{T}_{s}$ in Figs.~\ref{fig:fig3a}(b), \ref{fig:fig3a}(d) and \ref{fig:fig3a}(f), respectively. We consider three system sizes: $V=20^3$, $24^3$ and $V=28^3$. The behavior of $a$ is consistent with our previous observations, and its value appears independent of system size, except for $\hat{T}_{s}$ in the localized regime, though this may improve for larger $V$. In particular, the minimum of $-a$ is located close to $W_2^*\approx 16.5$. As discussed in the main text, the exponent $a$ for the operator $\hat{n}$ is related to the fractal dimension $d_2$ through $a=1-d_2\approx0.58$. The minimal value of $-a$ for this operator is in very good agreement with the above prediction, see the solid horizontal line in Fig.~\ref{fig:fig3a}(b). In contrast, the minimal values for the other operators deviate more substantially, see the solid horizontal lines in Figs.~\ref{fig:fig3a}(d) and \ref{fig:fig3a}(f).

\section{Mott pairs in localized phase}

\label{app:mott}

The differences between the behavior of typical and average fidelity susceptibilities for large disorder strengths can be understood as follows. (While differing in details, the argument presented below builds on the idea introduced in Ref.~\cite{Skvortsov_2022}.) Consider performing the perturbation theory in the localized regime, treating the hopping term in the Hamiltonian from Eq.~\eqref{eq_HA} as a perturbation. Within the forward scattering approximation~\cite{Lecture_notes}, the wave function amplitude, $\Psi_{n(b^*)}(b) = \langle b|n(b^*)\rangle$, can be written as
\begin{equation}
\label{eq:amplitude}
\begin{split}
    |\Psi_{n(b^*)} (b)| & \sim \left|\sum_{\pi:b^*\rightarrow b}\prod_{s=1}^{n}\frac{-1}{\varepsilon_{b^*}-\varepsilon_{\pi(s)}}\right|\sim \frac{1}{|\varepsilon_{b^*}-\varepsilon_b|}\left(\frac{1}{W}\right)^{n-1}\\
    & \sim \frac{1}{|\varepsilon_{b^*}-\varepsilon_b|}e^{-d(b^*,b)/\zeta}\,,
\end{split}
\end{equation}
for $b^*\neq b$ and $\Psi_{n(b^*)} (b^*)\sim 1$. In the above equation, $\ket{n(b^*)}$ denotes the $n$th single-particle energy eigenstate with the localization center at site $b^*$, $\varepsilon_{b^*}$ is the on-site potential, and $\pi$ is the set of the shortest trajectories connecting sites $b^*$ and $b$, each consisting of the minimal number of steps $n$. Furthermore, $d(b^*,b)$ is the Euclidean distance between sites $b^*$ and $b$, while $\zeta$ is the localization length. The expression in Eq.~\eqref{eq:amplitude} neglects possible resonances between sites $b^*$ and $\pi(s)$, which we address later.

We now focus on a single operator $\hat{n}$ defined in Eq.~\eqref{eq_randomized}, but the reasoning can be easily extended to other local operators, such as $\hat{T}$ or $\hat{T}_s$. Its off-diagonal matrix elements, which enter typical and average fidelity susceptibilities, can be calculated as
\begin{equation}
\begin{split}
    \langle n(b^*)|\hat{n}|m(c^*)\rangle & =\sum_{s=1}^{V} r_s \Psi^*_{n(b^*)}(s)\Psi_{m(c^*)}(s)\\
    & \sim e^{-d(b^*,c^*)/\zeta}\sum_{s=1}^{V} \frac{r_s}{(\varepsilon_{b^*}-\varepsilon_s)(\varepsilon_{c^*}-\varepsilon_s)},
\end{split}
\end{equation}
where $r_s$ is a i.i.d.~random number, as defined in Eq.~\eqref{eq_randomized} from the main text.
Single-particle energy eigenstates with nearby energies, $\omega_\text{typ} \le |\epsilon_{n(b^*)} - \epsilon_{m(c^*)}| \ll \Delta\epsilon$, have localization centers that are typically separated by large distances, $d(b^*,c^*)\sim V$, as discussed in the next paragraph. This spatial separation leads to an exponential suppression of $\langle n(b^*)|\hat{n}|m(c^*)\rangle$, explaining the trend of the rescaled $\chi_\text{typ}^r$ deep in the localized regime.

As clear from Eq.~\eqref{eq:amplitude}, when $e^{-d(b^*,c^*)/\zeta}\gg|\varepsilon_{b^*}-\varepsilon_{c^*}|$, the sites $b^*$ and $c^*$ are in resonance and the eigenstates $\ket{n(b^*)}$ and $\ket{m(c^*)}$ hybridize to form the so-called Mott pair~\cite{Mott01061968,Ivanov_2012}. Specifically, the Hamiltonian restricted to the two-dimensional Hilbert space spanned by $\ket{b^*}$ and $\ket{c^*}$ can be written as
\begin{equation}
\label{eq:twobody}
    \hat{H}_\text{2}=
    \begin{pmatrix}
\varepsilon_{b^*} & t_{b^*c^*} \\
t_{b^*c^*} & \varepsilon_{c^*}
\end{pmatrix},
\end{equation}
where we introduced $t_{b^*c^*}=e^{-d(b^*,c^*)/\zeta}$. Its eigenvalues are $\epsilon_{\pm}=\overline{\varepsilon}\pm\sqrt{\delta\varepsilon^2+t_{b^*c^*}^2}$, where $\overline{\varepsilon}=(\varepsilon_{b^*}+\varepsilon_{c^*})/2$ and $\delta\varepsilon=(\varepsilon_{b^*}-\varepsilon_{c^*})/2$. When sites $b^*$ and $c^*$ are in resonance, i.e., $|t_{b^*c^*}|\gg \delta\varepsilon$, its eigenstates become bonding and anti-bonding states separated by a gap $\Delta_\pm=\varepsilon_+-\varepsilon_-\approx|t_{b^*c^*}|$. They form a Mott pair, i.e., $\ket{\pm} =\left( \ket{n(b^*)} \pm \ket{m(c^*)} \right)/\sqrt{2}$. We note that such resonances are rare deep in the localized regime and occur for $d(b^*,c^*) \sim L$. Otherwise, level repulsion would be observed.

%%%FIGURE
\begin{figure}[!t]
\includegraphics[width=\columnwidth]{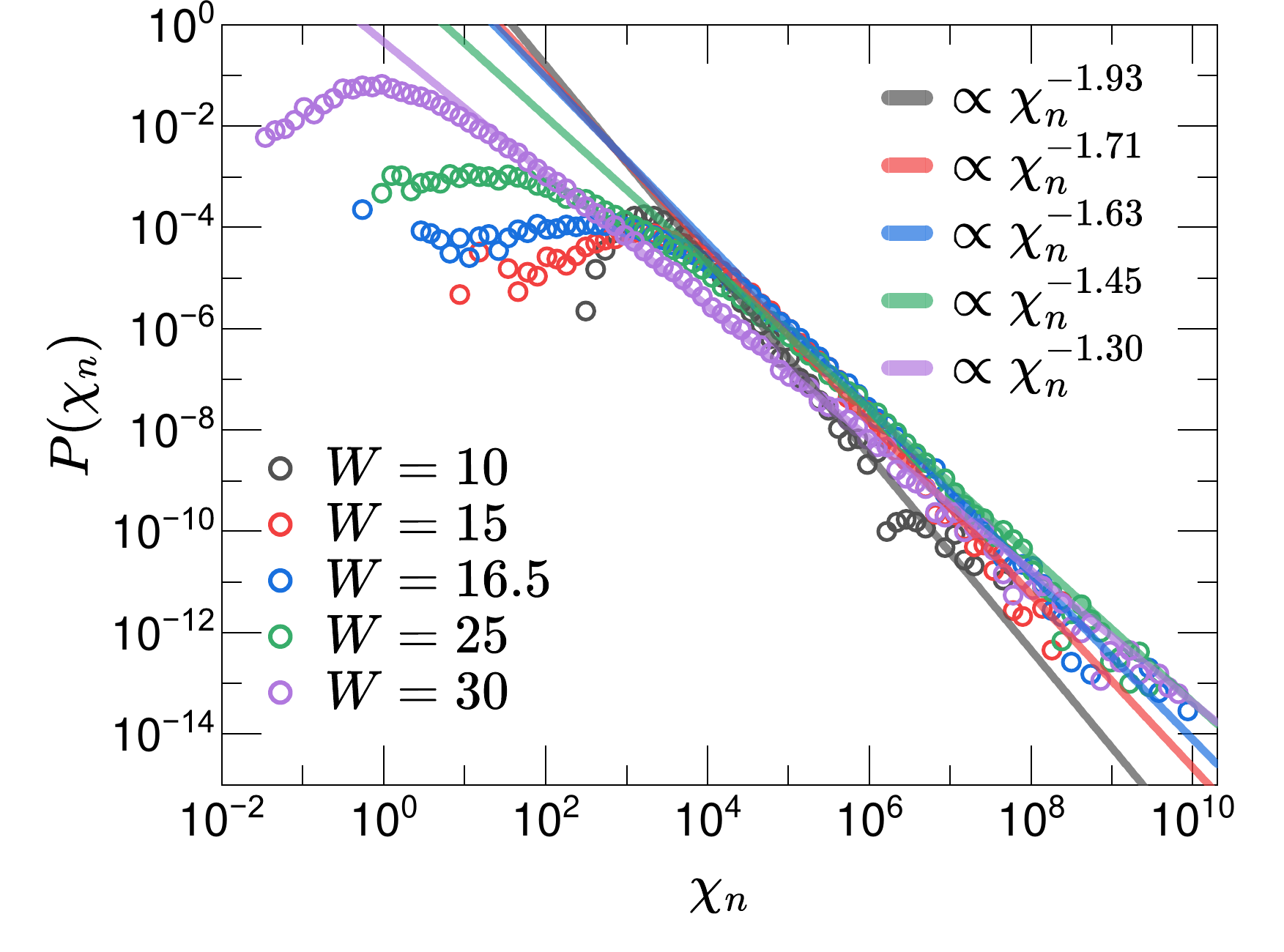}
\vspace{-0.2cm}
\caption{Distribution of fidelity susceptibilities, $P(\chi_n)$. We consider a single system size, $V = 28^3$, and several disorder strengths, $10 \le W \le 30$. The tails of distributions were fitted with a polynomially decaying function, $P(\chi_n) \sim \chi_n^{-p}$. These least-squares fits are shown as solid lines, and the corresponding exponents $p$ are indicated in the legend.
}
\label{fig:fig9}
\end{figure}
%%%FIGURE

%%%FIGURE
\begin{figure*}[t!]
\includegraphics[width=\textwidth]{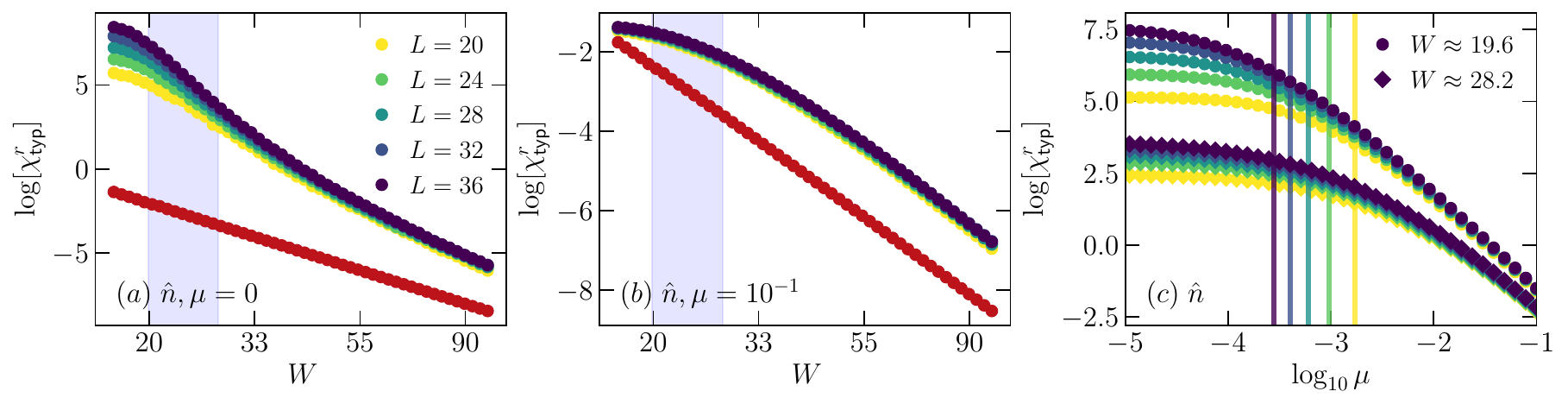}
\vspace{-0.2cm}
\caption{$\log[\chi_\text{typ}^r]$ plotted versus (a,b) $W$ and (c) $\log_{10}\mu$.
The perturbation is $\hat{n}$. In (a) and (b), the frequency
cutoffs are $\mu = 0$ and $\mu = 10^{-1}$, respectively. The upper curves show numerical results for the Anderson model, while the lower curve shows
the perturbation-theory prediction. The shaded region indicates the interval $W \in [20, 28]$, where $\tilde{\chi}_{\mathrm{typ}}^{\,r}$ exhibits a trend reversal. The horizontal axis is in logarithmic scale with tick marks at $\exp(3 + 0.5\, i)$,
where $i$ is an integer. Additionally, two disorder strengths are shown in (c): $W \approx 19.6$ and $W \approx 28.2$. The vertical line marks $\omega_{\mathrm{typ}}$.
}
\label{fig:fig4a}
\end{figure*}
%%%FIGURE

The off-diagonal matrix element of interest is given by
\begin{equation}
\begin{split}
        \bra{+}\hat{n}\ket{-}
        & = \sum_{s=1}^{V}\frac{r_s}{2}\left(|\Psi_{n(b^*)}(s)|^2+|\Psi_{m(c^*)}(s)|^2\right)\\
        & \sim \frac{r_{b^*}+r_{c^*}}{2},
\end{split}
\end{equation}
where we have taken into account that the wavefunctions are real. Importantly, this off-diagonal matrix element does not depend on the system size.

Since Mott pairs are rare deep in the localized regime, they are not expected to dominate $\tilde{\chi}_\text{typ}^r$, but they can significantly contribute to $\tilde{\chi}_\text{av}^r$. We highlight that the relevant contribution comes exclusively from pairs with energy gaps $\Delta_{\pm} \approx \mu$. We also note that the number of Mott pairs scales with the area of a sphere of radius $r \sim \zeta \log\Delta^{-1}_{\pm}$~\cite{Skvortsov_2022}. This explains why $\tilde{\chi}_\text{typ}^r$ and $\tilde{\chi}_\text{av}^r$ approach each other when $V$ is fixed and $\mu$ increases. Moreover, when the frequency cutoff $\mu$ is selected as a function decreasing with the system size $V$, the number of Mott pairs (as the number of all states) increases with $V$. In such a case, the distinction between the typical and average fidelity susceptibilities in the localized regime may persist even in the large system size limit.

\section{Fidelity susceptibility in localized phase}
\label{sec:pbc}

In Fig.~\ref{fig:fig9}, we complement the analysis of the differences between the average and typical fidelity susceptibilities from Sec.~\ref{sec:av_typical}.
Figure~\ref{fig:fig9} shows that the distribution of fidelity susceptibilities $\chi_n=\chi_n^r(\mu=0)$ (not regularized and not rescaled with $\mu$), for the sublattice kinetic energy $\hat{T}_{s}$, broadens with the increasing disorder strength $W$ close to and above the Anderson localization transition. This distribution was calculated for a single system size, $V = 28^3$, and several disorder strengths, $10 \le W \le 30$. It is natural to associate the increasing difference between the average and typical values with the developing tail of the distribution, so that while more fidelity susceptibilities $\chi_n$ become close to zero, larger fidelity susceptibilities also emerge. We fit this tail with a polynomially decaying function, $P(\chi_n) \sim \chi_n^{-p}$. The least-squares fits are shown as solid lines, while the corresponding exponents $\zeta$ are indicated in the legend.

In Fig.~\ref{fig:fig4a}, we show the same analysis as in Fig.~\ref{fig:fig10} from the main text, but for a different operator, i.e., the randomized site occupation $\hat{n}$. The results are qualitatively similar. Nevertheless, it is worth to emphasize that the prediction of the perturbation theory is different. This is due to the fact that $\langle n^{(0)}|\hat{n}|m^{(0)}\rangle=\delta_{nm}r_n$, so the offdiagonal matrix elements in the zeroth-order energy eigenstates vanish. As a result, the lowest-order approximation of the fidelity susceptibility of a single eigenstate $\ket{n}$, regularized but not rescaled, is given by
\begin{equation}
    \chi_n^r\approx\sum_{m\neq n} (r_n-r_m)^2\frac{|\langle n^{(0)}|\hat{T}|m^{(0)}\rangle|^2}{
        \left(\left[\omega_{n m}^{(0)}\right]^2 + \mu^2\right)^2},
\end{equation}
where $\hat{T}$ is the kinetic energy, $\langle n^{(0)}|\hat{T}|m^{(0)}\rangle=\delta_{mn'}$ and 
$n'$ are indices of nearest neighbors to $n$. Finally, $\chi_n^r$ is expected to scale as $1/W^4$ in the limit of small $\mu$.

\FloatBarrier
\bibliographystyle{biblev1}
\bibliography{references}

\end{document}